\documentclass[apj]{emulateapj}
\usepackage[latin1]{inputenc}
\usepackage{amsmath}
\usepackage{graphics}

\usepackage[usenames,dvipsnames]{color}

\newcommand{\sigT}[1]{\mbox{$\sigma_{\mbox{\scriptsize T}}^{#1}$}}
\newcommand{\sigSB}[1]{\mbox{$\sigma_{\mbox{\scriptsize SB}}^{#1}$}}
\newcommand{\kboltz}[1]{\mbox{$k_{\mbox{\scriptsize B}}^{#1}$}}
\newcommand{\MdotE}[1]{\mbox{$\dot{M}_{\mbox{\scriptsize Edd}}^{#1}$}}
\newcommand{\LEdd}[1]{\mbox{$L_{\mbox{\scriptsize Edd}}^{#1}$}}

\newcommand{\Rcirc}[1]{\mbox{$R_{\mbox{\scriptsize circ}}^{#1}$}}
\newcommand{\Ric}[1]{\mbox{$R_{\mbox{\scriptsize iC}}^{#1}$}}
\newcommand{\Mbh}[1]{\mbox{$M_{\mbox{\scriptsize bh}}^{#1}$}}
	
\newcommand{\Msol}[1]{\mbox{$M_{\odot}^{#1}$}}
\newcommand{\Rsol}[1]{\mbox{$R_{\odot}^{#1}$}}
\newcommand{\Lsol}[1]{\mbox{$L_{\odot}^{#1}$}}
\newcommand{\Mdot}[1]{\mbox{$\dot{M}^{#1}$}}
	
\newcommand{\mdot}[1]{\mbox{$\dot{m}^{#1}$}}
\newcommand{\Mtot}[1]{\mbox{$\Mdot{#1}_{\mbox{\scriptsize tot}}$}}
\newcommand{\Mddot}[1]{\mbox{$\dot{M}_{d}^{#1}$}}
	\newcommand{\innrMddot}[1]{\mbox{$\dot{\mathfrak{M}}_{d}^{#1}$}}
\newcommand{\Mcdot}[1]{\mbox{$\dot{M}_{c}^{#1}$}}
	\newcommand{\innrMcdot}[1]{\mbox{$\dot{\mathfrak{M}}_{c}^{#1}$}}
	\newcommand{\mcdot}[1]{\mbox{$\dot{m}_{c}^{#1}$}}

\newcommand{\Mzdot}[1]{\mbox{$\dot{M}_{z}^{#1}$}}
\newcommand{\Msdot}[1]{\mbox{$\dot{M}_s^{#1}$}}
	
\newcommand{\Mwdot}[1]{\mbox{$\dot{M}_w^{#1}$}}

\newcommand{\Lcrit}[1]{\mbox{$L_{\mbox{\scriptsize cr}}^{#1}$}}

\newcommand{\Psys}[1]{\mbox{$P_{\mbox{\scriptsize sys}}^{#1}$}}
\newcommand{\dsys}[1]{\mbox{$d_{\mbox{\scriptsize sys}}^{#1}$}}

\newcommand{\Tbb}[1]{\mbox{$T_{\mbox{\scriptsize bb}}^{#1}$}}

\newcommand{\fcol}[1]{\mbox{$f_{\mbox{\scriptsize col}}^{#1}$}}
\newcommand{\rhoL}[1]{\mbox{$\rho_{\mbox{\scriptsize L1}}^{#1}$}}
\newcommand{\HsubL}[1]{\mbox{$H_{\mbox{\scriptsize L1}}^{#1}$}}

\begin{document}
\title{Prototyping non-equilibrium viscous-timescale accretion theory using LMC X-3}

\author{Hal J. Cambier}
\affil{Physics Department, University of California, Santa Cruz, CA 95064}

\and

\author{David M. Smith}
\affil{Physics Department, University of California, Santa Cruz, CA 95064, USA}
\affil{Santa Cruz Institute for Particle Physics, University of California, Santa Cruz, CA 95064, USA}

\begin{abstract}
	
Explaining variability observed in the accretion flows of black hole X-ray binary systems remains challenging,
especially concerning timescales less than, or comparable to, the viscous timescale but much larger than the inner
orbital period despite decades of research identifying numerous relevant physical mechanisms. We take a simplified
but broad approach to study several mechanisms likely relevant to patterns of variability observed in the persistently
high-soft Roche-lobe overflow system LMC X-3. Based on simple estimates and upper bounds, we find that physics
beyond varying disk/corona bifurcation at the disk edge, Compton-heated winds, modulation of total supply rate
via irradiation of the companion, and the likely extent of the partial hydrogen ionization instability is needed to
explain the degree, and especially the pattern, of variability in LMC X-3 largely due to viscous dampening. We then
show how evaporation--condensation may resolve or compound the problem given the uncertainties associated with
this complex mechanism and our current implementation. We briefly mention our plans to resolve the question,
refine and extend our model, and alternatives we have not yet explored.
	
\end{abstract}

	\section{Introduction}
	\label{sec:Intro}

\begin{figure}
	\centering
	\includegraphics[width=.48\textwidth]{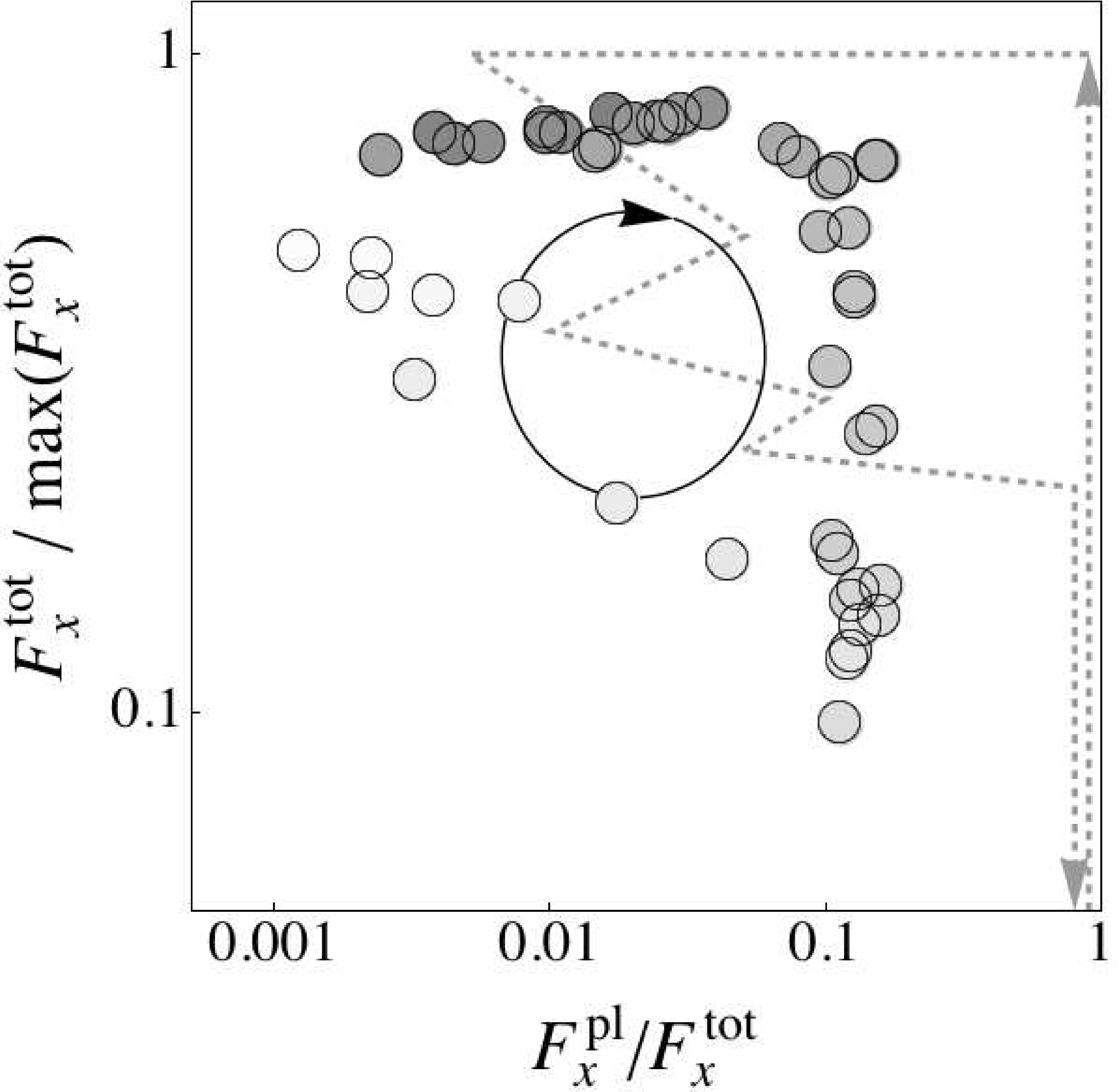}
	\caption{Shaded points show a typical spectral evolution cycle of LMC X-3 (first episode in fig.\ref{fig:MdotsData}) with darker points corresponding to earlier observations, showing that it winds around in the opposite sense of typical transients' cycles shown schematically in the gray, dashed ``q'' or ``turtle-head''.}
	\label{fig:Hysteresis}
	\end{figure}

	The X-ray spectrum of black hole X-ray binaries (BHXRBs) often shows variability in intensity and hardness on timescales of order the viscous timescale, but much larger than the innermost orbital period, including the well-known ``q''-diagram hysteresis patterns traced by transient BHXRBs \citep[][and see fig.\ref{fig:Hysteresis}]{Fender_etal_1999_GX339_jetQuench, HomanBelloni_05_qHystRev, DoneGierKub_07Rev}.  The properties of such variability may also evolve over multiple viscous-timescale cycles.  The high-soft (but sub-Eddington) and quiescent intensity-hardness limits are understood as manifestations of the thin-disk \citep{ShakSun73} and radiatively-inefficient advection-dominated accretion flow (ADAF; \citeauthor{NaYi95a} \citeyear{NaYi95a}) accretion limits while states between are typically understood as some evolving combination of disk and ADAF flows \citep{ChakrabartiTitarchuk_1995_TwoComp,Esin_etal_1997_TwoComp,Nandi_etal_2012_TwoFlow}.  Theoretical work has begun in this regime, but still cannot fully explain observations, especially regarding viscous timescales where detailed simulations require prohibitively many time steps and steady-state assumptions lose validity.  This motivates us to develop theory in more detail starting with a system like LMC X-3, whose behavior is more constrained than that of the transients, but still exhibits substantial variability that is quantitatively challenging to explain. 
	
	In the current paradigm, transient BHXRBs are those systems where the outer disk reaches temperatures low enough to trigger the partial hydrogen ionization instability (PHII) in which accretion proceeds via cycles as viscosity alternately concentrates mass into rings and diffuses it inward (see \citeauthor{Cannizzo_98_IoniInstCycles1} \citeyear{Cannizzo_98_IoniInstCycles1}, and \citeauthor{Lasota_2001_IoniInstRev} \citeyear{Lasota_2001_IoniInstRev} for a review).  The part of the cycle where viscosity concentrates mass leads to the quiescent phase where any emission is presumably powered by some fraction of the flow that escapes the viscosity trap.  Eventually, density increases enough to raise disk temperature above the transition point in viscosity, and the sudden jump in accretion rate leads to a rise in luminosity (often several orders of magnitude) while still in the hard state, followed by a softening of the spectrum at this same, peak luminosity (the vertical and horizontal shifts by the dashed line in fig.\ref{fig:Hysteresis}).  Jet emission is seen to shut off as the system enters the extreme high-soft state \citep{Fender_etal_1999_GX339_jetQuench}.  Afterward, the system typically makes partial transitions in hardness (with small, erratic shifts in luminosity) on sub-viscous timescales and associated with intermittent jet emission on top of a secular decline in luminosity.  Eventually the system transitions completely back to the hard state and then finishes fading back into quiescence.
		
	There are a handful of persistent BHXRBs and black-hole-candidate X-ray binaries that do not execute such extreme quiescence-flaring cycles, and consistent with the transient paradigm they appear to avoid the PHII through an appropriate combination of disk size, luminosity (for disk irradiation), and mass supply rate \citep{Coriat_etal_12_PersistentVTransients}.  Despite the label, these systems may still show significant variability and a variety of behaviors.  Cygnus X-1 appears to slide between canonical high-soft and low-hard states with some scatter but no clear hysteresis \citep[hereafter SHS02]{SmHeSw02}.  The black hole candidates GRS 1758--258 and 1E 1740.7--2942 exhibit transient-like hysteresis but seldom reach the soft or quiescent limits (SHS02).  LMC X-3 is typically bright in soft X-rays, but shows some hysteresis that circulates in the opposite direction to transients (see fig.\ref{fig:Hysteresis}), and does occasionally transition completely to the hard state \citep{Wilms_etal_2001_LMCXsObs,SmaleBoyd_2012_arXiv}.
	
\begin{figure*}[!ht]
	\centering
	\includegraphics[width=0.60\textwidth]{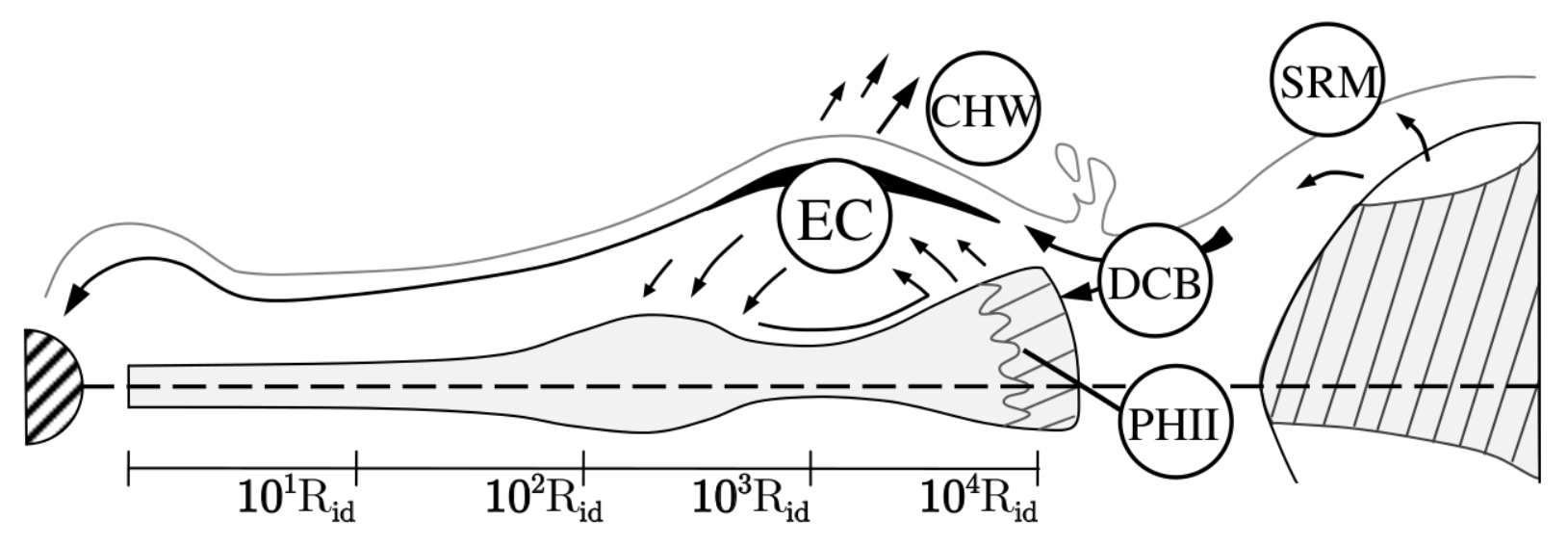}
	\caption{A cartoon of the primary mechanisms considered for driving long timescale changes observed in inner disk and corona accretion rates.  These are: supply-rate modulation (SRM, \S\ref{sec:mechs_SRM}) from the companion, disk-corona bifurcation (DCB, \S\ref{sec:mechs_DCB}) at the disk's edge, where disk warping may affect SRM and DCB, the partial hydrogen ionization instability (PHII, \S\ref{sec:mechs_PHII}), Compton-heated winds (CHW, \S\ref{sec:mechs_CHW}) at large radii, and evaporation and condensation (EC, \S\ref{sec:mechs_EC}) exchanging mass between disk and corona.}
	\label{fig:BasicMechs}
	\end{figure*}

	Accounting for irradiation of the outer disk, LMC X-3 accretes at rates high enough to usually avoid the ionization instability completely \citep{Coriat_etal_12_PersistentVTransients}, though the outer disk likely becomes susceptible for the deeper drops in disk luminosity, and almost certainly for rare, complete state transitions.  Such high accretion rates are explained via Roche-lobe overflow (RLO); optical measurements of the companion's spectral type that take disk irradiation into account indicate B5IV \citep{Soria_etal2001} or B5V \citep{Va_BaNoNe_2007_LMCX3SysProps} spectral type.  Such a star will fill the Roche lobe for a 1.7 day orbit around a black hole with mass equal to the recent 9.5$\Msol{}$ lower bound \citep{Va_BaNoNe_2007_LMCX3SysProps}.  Furthermore, \citeauthor{Soria_etal2001} \citeyearpar{Soria_etal2001} have argued that feeding the observed X-ray luminosities via winds would lead to column densities far higher than measured.  Although its distance precludes direct measurement or exclusion of radio jets, the jet quenching observed in transient BHXRBs as they approach the high-soft state also suggests that LMC X-3 does not usually possess strong jets \citep{FenderSoTz_1998_jetlimits}.  Thus, if LMC X-3 shares similar variability mechanisms with transients, while being far less prone to the ionization instability and jet outflows, then LMC X-3 provides a more controlled setting to study such mechanisms.  Below we list the major mechanisms we have considered so far in modeling LMC X-3, also summarized in fig.\ref{fig:BasicMechs}.
	
	For RLO systems like LMC X-3, hard X-rays from the inner accretion disk can lead to supply-rate modulation (SRM) from the companion by inflating the companion's atmosphere to increase the density of gas at the L1 Lagrange point, as well as the area and pressure behind the nozzle \citep{LubShu75,MeMe_Ho83}.  We elaborate in \S\ref{sec:mechs_SRM}.
	
	As the stream of gas from the companion dissipates energy and spirals onto the circularization radius, it may encounter the edge of the viscously spreading disk and from this point the flow can undergo disk corona bifurcation (DCB) as some fraction can efficiently shock, cool, and join the disk while some fraction may stream past the thin disk edge and maintain its virial temperature \citep{Hessman99,ArmLiv98}.
	
	Warping of the outer disk, whether driven by irradiation \citep{Pringle_1992_BasicWarping,OgDu01,Foulkes_etal_2010_IrradWarping} or a lift force \citep{MontgomeryMartin_2010_Lift} can affect the accretion flow by changing the density profile that the disk presents to the RLO stream (\S\ref{sec:mechs_DCB}), and by varying how the companion is exposed to or shadowed from inner disk X-rays.  For LMC X-3 specifically, \citeauthor{OgDu01} (\citeyear{OgDu01}) and \cite{Foulkes_etal_2010_IrradWarping} predict that the system is potentially unstable to irradiation-driven warping.  Because the dominant long-term effects of warping are through bifurcation, and because we will lump them together as a boundary condition in our simulations, we combine discussion of them into \S\ref{sec:mechs_DCB}.	
	
	X-rays from the inner disk can Compton heat gas in the disk atmosphere and any corona at large radii above the local virial temperature thus driving a Compton-heated wind (CHW), discussed in \S\ref{sec:mechs_CHW}, which is not only important for removing gas, but for removing hot corona that might otherwise help evaporate the disk or condense further inward (item (EC) below).  As noted, for large enough disks and insufficient irradiation, a finite strip in the outer disk becomes susceptible to the PHII.  For now, we do not treat it in any detail, but discuss the likely extent and manner of its effects in LMC X-3 in \S\ref{sec:mechs_PHII}.
		
	If the disk and corona are coupled thermally, then the disk and corona may exchange mass through evaporation and condensation (EC).  \citeauthor{MaPr07} (\citeyearpar{MaPr07}, hereafter MP07) provide a thorough introduction and numerical treatment, and \citeauthor{LiTaMe_HoMe07} (\citeauthor{LiTaMe_HoMe07}, hereafter LTMHM07) and \cite{Me_HoLiMe09} discuss more applications and provide the steady-state prescription for our modified method.  Through EC, extant disks will tend to preserve the soft state down to lower luminosities via Compton-cooling-driven condensation, providing a natural explanation of why BHXRBs return to the hard state at lower luminosity and thus show hysteresis (\citeauthor{Me_HoLiMe09} \citeyear{Me_HoLiMe09} focus on this aspect).  We discuss other interesting effects possible in \S\ref{sec:mechs_EC}.
	
	We will restrict our focus to an alpha-viscosity prescription for the disk.  For the present work describing long-timescale variability over accretion rates typical to LMC X-3 where uncertainties regarding conventional mechanisms still loom large, we consider this perfectly adequate, but acknowledge the possibility of more intrinsic variability mechanisms (\S\ref{sec:DiscConc}).
	
	In \S\ref{sec:LMCX3} we review key features of LMC X-3's accretion behavior, summarize how we infer the innermost disk and corona/ADAF accretion rates from the X-ray data (additional details are provided in the Appendix), and critically examine the qualitatively simple bifurcation-only model in \citeauthor{SmDaSw07} (\citeyear{SDS07}, hereafter SDS07).  The latter motivates \S\ref{sec:mechs}, in which we furnish additional detail on the mechanisms as listed above, including estimated constraints on their effects and their current level of implementation in our modeling.  In \S\ref{sec:Res_noEC} we argue that a model including mechanisms besides EC cannot explain the data, but does best when given an unreasonably small disk radius and very large variations in corona-disk ratio at this boundary.  We then show in \S\ref{sec:Res_EC} how EC may effectively recreate such seemingly ad-hoc conditions, but how it may also imply behavior inconsistent with observations, including an extremely easily triggered ``sympathetic'' mode where the innermost disk and corona accretion rates rise and fall simultaneously.  We briefly review the results and caveats of the current model, and state our current plans to resolve the question in \S\ref{sec:DiscConc}.
	
	\section{LMC X-3 as prototype}
	\label{sec:LMCX3}
	
\begin{figure}[ht!]
	\centering
	\includegraphics[width=.5\textwidth]{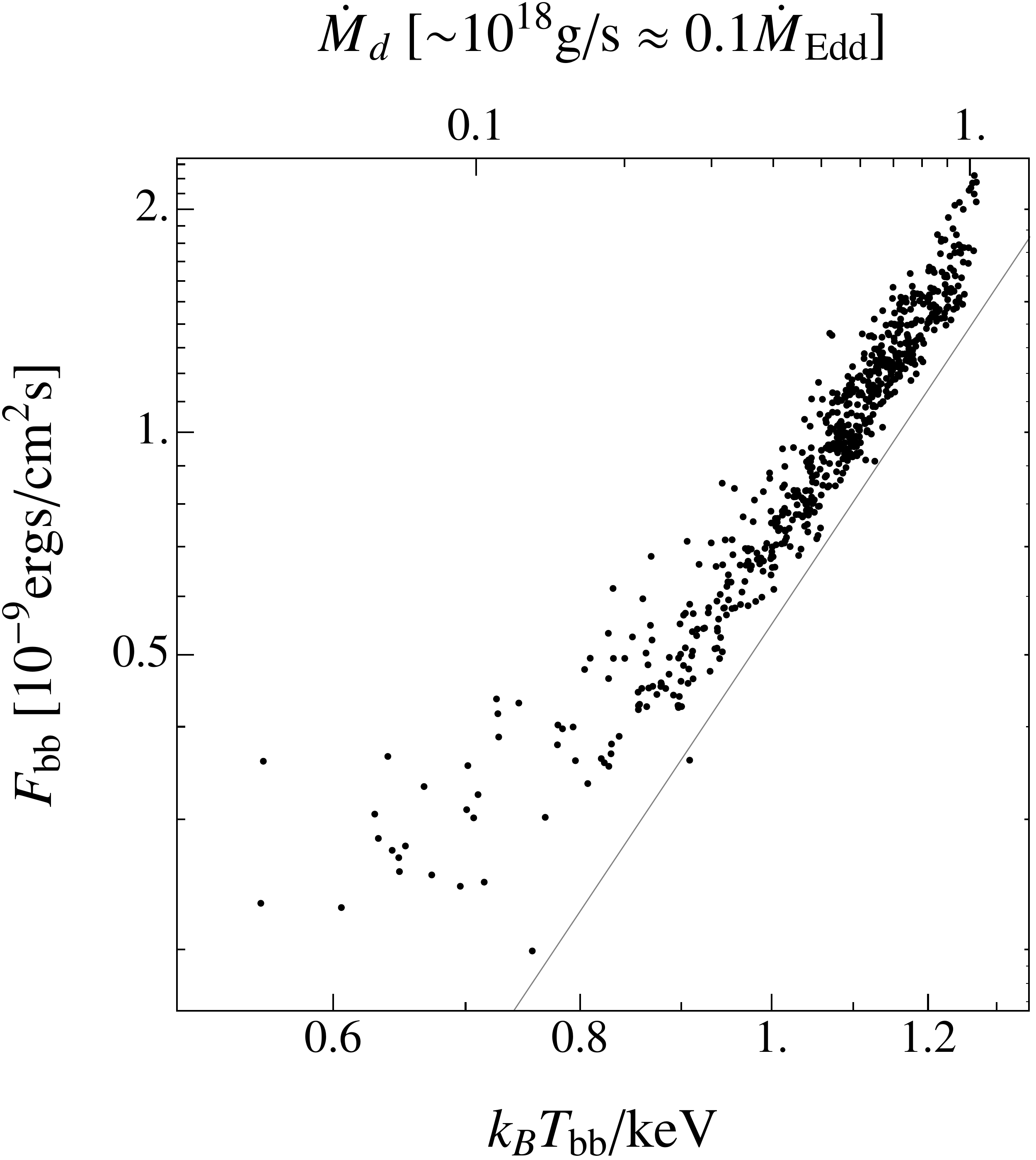}
	\caption{Fitted blackbody-component fluxes plotted against fitted temperature with a pure $T^{4}$ curve in faint gray for comparison.}
	\label{fig:FluxTemp}
	\end{figure}

	Besides simplifying initial modeling as discussed above, LMC X-3 offers additional practical advantages.  The \textit{Rossi X-ray Timing Explorer} (\textit{RXTE}) monitored LMC X-3 for over 16 years, and at least five of those include observations each about a kilosecond long taken roughly twice a week, thus providing a long, uninterrupted history of accretion with sufficient resolution at the timescales we seek to study.  Also, the X-ray blackbody component, when present, tracks the Stefan--Boltzmann law fairly well (see fig.\ref{fig:FluxTemp}) indicating that inner disk geometry (i.e. truncation, warping) changes fairly little, and that the corona optical depth, $\tau_c$, is small, simplifying estimates of the inner corona accretion rate, $\innrMcdot{}$.  Unless specified otherwise,  we will use symbols $\innrMddot{}$ ($\innrMcdot{}$) as shorthand for disk (corona) accretion rates at the inner disk radius, $R_{id}$, and reserve the italic-face for general, local accretion rates $\Mddot{}=\Mddot{}(R,t)$, $\Mcdot{}=\Mcdot{}(R,t)$.  Also, unless otherwise indicated, we will use the following system parameters: black hole mass, $\Mbh{}$=$10\Msol{}$, companion mass $M_*$=$5\Msol{}$, orbital period $\Psys{}$=$1.705$d, inclination $i$=$67^{\rm o}$ and system distance, $\dsys{}$=$48$ kpc \citep{vdK_etal_1985, Va_BaNoNe_2007_LMCX3SysProps}, which also imply a circularization radius of $\Rcirc{}=2.7 \times 10^{11}$cm.
	
\begin{figure*}[ht]
	\includegraphics[width=1.\textwidth]{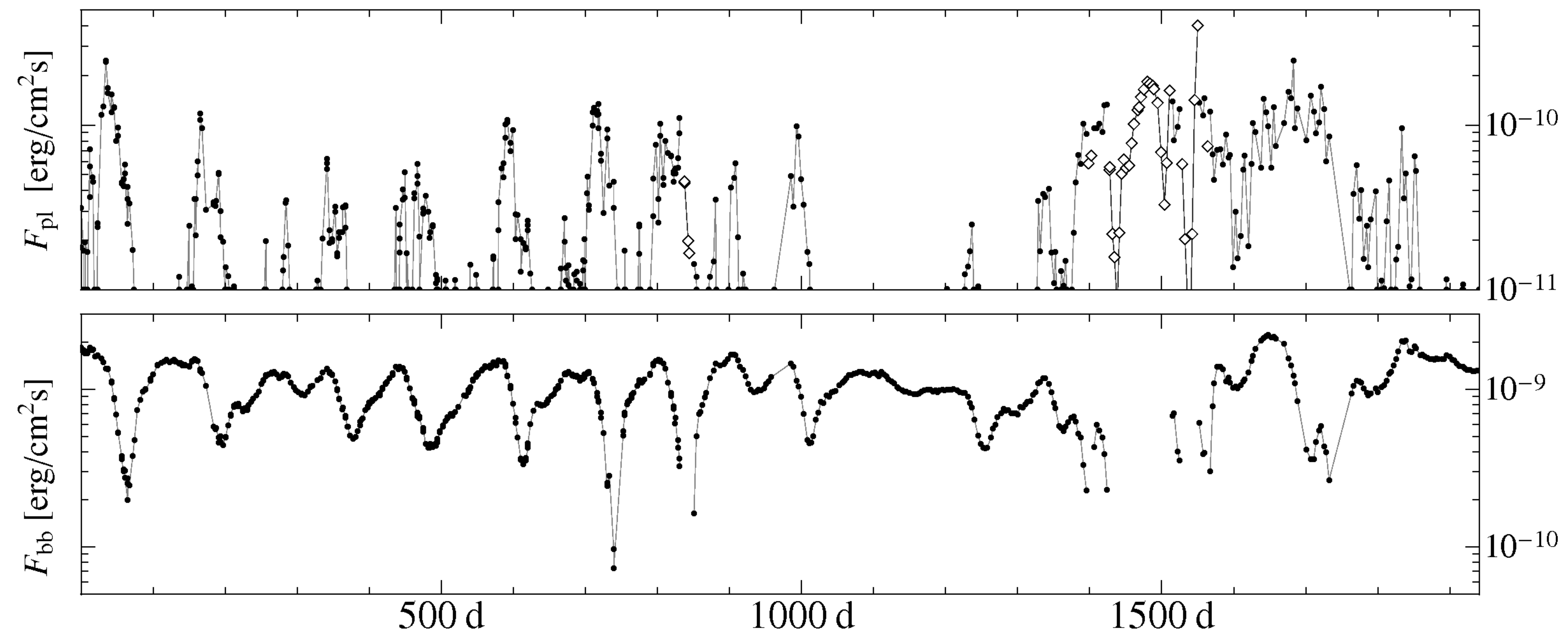}
	\caption{Fitted power-law (top panel) and blackbody (bottom panel) components of LMC X-3's X-ray flux since 53436.1 MJD.  Diamond points in the top panel mark observations categorized as the pure hard state by the criteria in \S\ref{sec:LMCX3}.}
	\label{fig:rawFluxes}
	\end{figure*}

	We first fit individual \textit{RXTE} spectra with a disk blackbody and a power law of fixed photon index, $\Gamma_{\mbox{\scriptsize pli}} =2.34$ (as in SDS07) with total absorption of fixed column density $n_H = 3.8\times 10^{20}$cm$^{-2}$ \citep{Page_etal_2003_LMCX3obs}, using the \texttt{wabs*simpl*diskbb} models in XSPEC \citep{Arn96}.  To systematically identify transitions to the low-hard state we looked for cases where the first fitting gave reduced $\chi^2>1.1$, and refit these with a \texttt{wabs*(plaw)} model where the power-law index is not frozen.  Reassuringly, spectra identified this way were fit better with fewer parameters, and are also typically preceded by obvious declines in the blackbody component (fig.\ref{fig:rawFluxes}).
		
\begin{figure*}[!ht]
	\includegraphics[width=1.\textwidth]{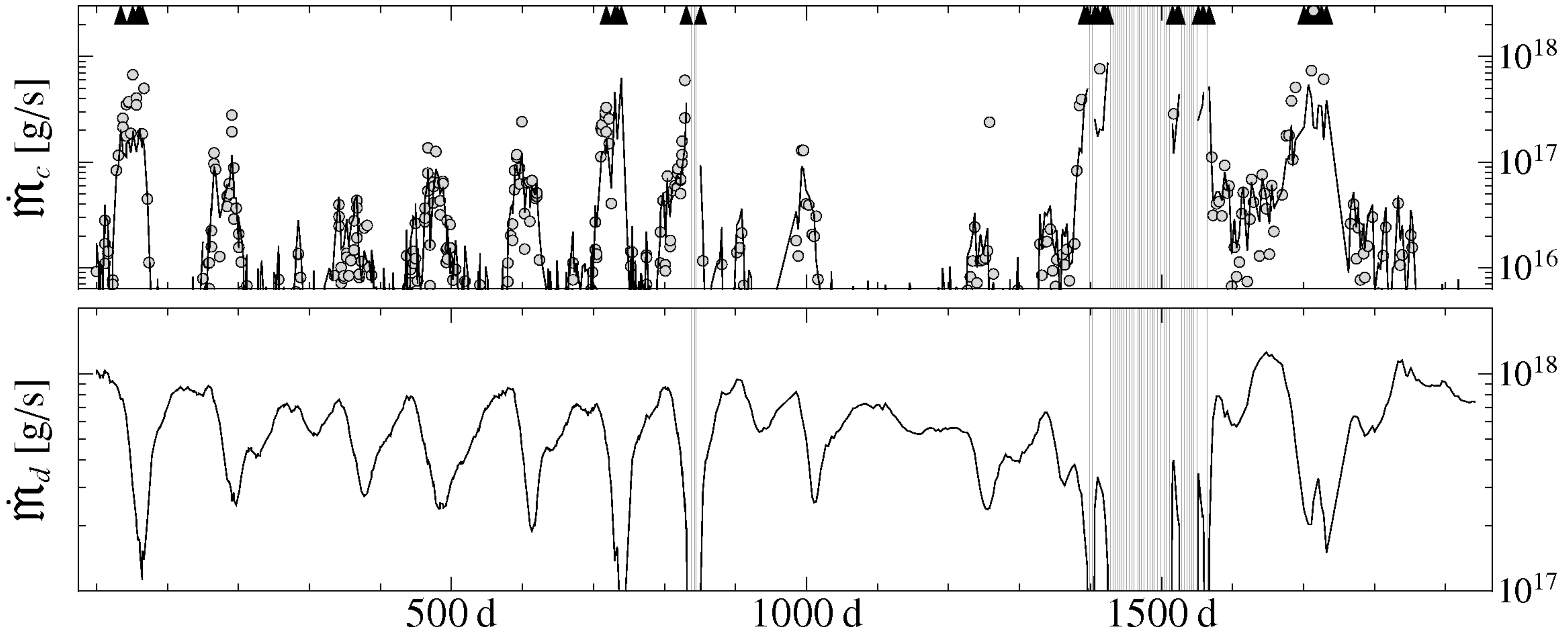}
	\caption{Inferred accretion history since 53436.1 MJD, barring times LMC X-3 was observed in the low-hard state (gray vertical lines) where inferring accretion rates is more ambiguous.  Note the vertical labels refer to inner accretion rates here.  In the top panel, solid trace shows simple, direct estimate of $\innrMcdot{}$ as well as results (points) of a more detailed method and arrowheads indicate points where the detailed method required abnormally high $\innrMcdot{}$ (see \S\ref{sec:Appendix_Norms}).  Overall, one can see trend for $\innrMcdot{}$ to pulse ``on'' quasi-periodically and anticipate episodic drops in $\innrMddot{}$}
	\label{fig:MdotsData}
	\end{figure*}

	For low $\tau_c$ one can describe the flows qualitatively by taking $\innrMddot{}(t)\sim \Tbb{4}$ and $\innrMcdot{}(t)$ proportional to the ratio of power-law to blackbody count fluxes (as in SDS07, though there the disk central temperature was confused with the effective temperature giving $\innrMddot{}(t)\sim \Tbb{20/6}$).  We obtain absolute normalization for $\innrMddot{}$ by fixing $R_{id}$ and comparing observed and predicted fluxes in the high state where agreement should be best, while for $\innrMcdot{}$, we obtain an estimate based on the simple $\tau_c$ and a typical ADAF solution, and check this against a more detailed calculation.  We relegate the details to the Appendix to focus on a general description of accretion behavior (fig.\ref{fig:MdotsData}).

	From fig.\ref{fig:MdotsData}, one can see that the $\innrMcdot{}$ ``turns on'' in pulses (referring to the secular month-long features and not the jagged week-long sub-pulses) roughly a viscous timescale apart and slightly shorter in duration, and that these pulses tend to anticipate drops in $\innrMddot{}$.  This trend was already noted in \citep{SmDaSw07} based on inferred qualitative accretion rates, and led the authors to posit a ``bifurcation-only'' model where a fairly-constant total supply rate ($\Msdot{}$) is split far from the black hole between non-interacting quickly-draining-corona and slowly-draining-disk components.  Our normalization estimates for $\innrMcdot{}(t)$ suggest that for any given episode there is generally insufficient total mass in a $\innrMcdot{}(t)$ pulse to explain the associated $\innrMddot{}$ drop.  Even if our overall normalization is off, we still found that scaling $\innrMcdot{}$ to conserve mass for one episode does not work very well for other episodes.  This mass-conservation problem motivated considering mechanisms that can adjust the total supply rate, remove mass, and/or exchange it between disk and corona flows.
		
	The secular evolution of the episodes on super-viscous timescales also lends itself to interpretation as multiple mechanisms acting on similar timescales effectively generating ``beat-frequency'' behaviors.  The simple alternative of some mechanism(s) acting on super-viscous timescales coupled to viscous-timescale variability mechanisms lacks good candidates for the former.  Nuclear evolution is too slow, we do not expect significant magnetic cycles from a companion with a radiative outer envelope (but keep the possibility in mind regarding other systems), and the inferred mass ratio in LMC X-3 is too high for slowly-growing tidal resonances to be significant \citep{FKR_book02}.  Furthermore, based on the observed inclination, the warps would have to reach heights of $30^{\rm o}$ relative to the orbital plane, and survive the severe drops in $\innrMddot{}$, to exhibit precession effects if irradiation-driven, which poses difficulties if LMC X-3 is only marginally unstable to irradiation-driven warping as \cite{OgDu01} suggest.
	
	The disk component in LMC X-3 tends to fall and recover more rapidly for larger drops than for shallow drops, a trend quantified in SDS07 and recently over an expanded data set in \citeauthor{SmaleBoyd_2012_arXiv} (\citeyear{SmaleBoyd_2012_arXiv}).  This aspect is qualitatively consistent with a bifurcation-only model given sufficient variation in the amplitude and duration of a drop at the outer edge - sensitivity to duration for a single input amplitude can be seen in figures 7\&8 of \citet{Zdz_etal_2009_visc}.  However, using their analytical machinery, with and without crude representation of SRM and outflow effects, we will later show that rough quantitative agreement with observations of LMC X-3 requires inputs that are extremely unlikely without additional physics (\S\ref{sec:Res_noEC}).  An interesting exception to the usual of $\innrMcdot{}$ pulses heralding steep $\innrMddot{}$ drops is the small drop in $\innrMddot{}$ at 1100d into fig.\ref{fig:MdotsData} not associated with any $\innrMcdot{}$ pulse above the typical noise level.

	Inferring accretion rates in the absence of the disk component introduces additional parameters and uncertainties, but we wish to make a few relevant observations while we work on a more definitive analysis of the hard state.  The disk component drops and recovers on timescales of days in transitions into and out of the hard state, and tends to return more quickly than it decays when LMC X-3 is at its ``hardest'' in our data, circa the 1500d mark in bottom of fig.\ref{fig:MdotsData}, consistent with the notion of an extant inner disk preserving itself through condensation.  Also, the power-law component tends to increase before failed and successful disk restarts, which may physically correspond to the inner edge of a truncated disk moving inward to provide more and hotter seed photons, and/or rapid condensation.
	
	\section{Variability mechanisms considered}
	\label{sec:mechs}

	Though the basic physics of companion irradiation and streaming are simple, the dynamics are potentially complicated to initialize and implement in detail, especially if the outer disk warps.  However, we can estimate bounds on both mechanisms individually, and because they sit at the edge of the accretion flows, we can lump them into a manual boundary condition for now and still derive meaningful results.  Compton-heated winds can be launched a bit further inward, but can be described fairly well by simple analytical functions of radius and X-ray luminosity assuming that the corona is easily replenished, and thus we can quickly obtain upper bounds on CHW losses.
	
	Evaporation--condensation can depend sensitively on disk and corona conditions at all radii making it the least amenable to simple estimates, and as noted earlier this same strong dependence on the system's state can naturally engender hysteresis.  EC also allows the disk component to vary more substantially and more rapidly by evaporating disk material interior to the circularization radius, but this evaporated disk material can also condense further inward much faster than inner disk conditions change, potentially to the point that $\innrMddot{}$ rises and falls simultaneously with $\innrMcdot{}$.  This ``sympathetic'' accretion mode can be seen in the more detailed simulations of Mayer and Pringle (their fig.8) and in many cases we simulated (e.g. figures \ref{fig:MP07_Comparison},\ref{fig:results_DfaultDeltaR},\ref{fig:ETB3comb}), but is effectively absent (or negligible) in our observations of LMC X-3, and thus primarily poses a challenge to our basic EC model.  %As noted, it may actually be manifest in LMC X-3 as the occasional bumps in $\innrMddot{}$ preceding declines, but against the much greater prevalence and strength of this effect in our simulations it primarily poses a challenge to our basic EC model (\S\ref{sec:Res_EC}).
	
		%\FloatBarrier
		\subsection{SRM Estimates and Remarks}
		\label{sec:mechs_SRM}
	
	The total supply rate of mass through the L1 nozzle $\Msdot{}$, will scale with the product of local gas density $\rhoL{}$, speed at which gas streams through the nozzle (roughly the local sound speed $c_s$), and area of the nozzle $A_n$ where the latter has width and height roughly equal to the isothermal scale height in the local tidal field, $\HsubL{2}\approx c_s^2/\Omega_{\mbox{\scriptsize orb}}^2$ \citep{LubShu75}.  %Near the L1 point, the density will fall with altitude above the photosphere, $d-d_{\mbox{\scriptsize ph}}$ like $\rhoL{}=\rho_0 \exp(-(d-d_{\mbox{\scriptsize ph}})^2/\HsubL{2})$ where $d$ is distance from the center of the companion and $\rho_0$ is density at the base of the atmosphere.  
	Under X-ray irradiation, each layer of the atmosphere will tend to heat up until it emits the intrinsic stellar flux plus the incident X-ray flux at that altitude.  Due to the very steep transition in density at the photosphere, we find most of the X-ray energy is deposited in a thin layer there, which we will take to be infinitesimally thin for now.  Thus, the modulation with respect to a given reference state as a function of stellar temperature $T_{\star}$, effective incident X-ray luminosity $L_{x,\mbox{\scriptsize eff}}$, gravity-darkened stellar luminosity $L_{\star,\mbox{\scriptsize eff}}$, and distance between L1 and the photosphere $d_{L1}-d_{\mbox{\scriptsize ph}}$ is given by \citep[e.g.][]{MeMe_Ho83}:

\begin{equation}
\centering
\label{eqn:MtotMod}
\dfrac{\Msdot{}}{\Msdot{\mbox{\scriptsize ref}}} = \left(\dfrac{T_{\star}}{T_{\star}^{\mbox{\scriptsize ref}}}\right)^{3/2} \exp\left[\left(\dfrac{d_{L1}-d_{\mbox{\scriptsize ph}}}{\HsubL{\mbox{\scriptsize ref}}}\right)^2 \dfrac{ T_{\star}^{\mbox{\scriptsize ref}} - T_{\star} }{ T_{\star}^{\mbox{\scriptsize ref}} }\right]
\end{equation}
where
\begin{equation}
\centering
\dfrac{T_{\star}}{T_{\star}^{\mbox{\scriptsize ref}}} = \left(\dfrac{L_{x,\mbox{\scriptsize eff}}+L_{\star,\mbox{\scriptsize eff}}}{L_{x,\mbox{\scriptsize eff}}^{\mbox{\scriptsize ref}}+L_{\star,\mbox{\scriptsize eff}}}\right)^{1/4}.
\end{equation}

	%Research on driving modulation at relevant timescales by irradiation (e.g. \cite{MeMe_Ho83, Osaki85, GoHa_1993_CompModEqns}), thus naturally focuses on cataclysmic variables and neutron-star low-mass X-ray binaries which realize large ratios of incident X-ray flux to intrinsic stellar flux.  
	Short of solving the structure of the stellar envelope in the Roche-lobe potential under time-varying irradiation, we can estimate the extent of SRM by computing the ratio of effective incident-to-intrinsic luminosity.  One can make a simple estimate by computing the effective gravity at a point sitting about halfway between the nozzle and the pole of the companion giving $L_{*,\mbox{\scriptsize eff}}/L_* \approx 0.68$, and also use the inclination of this point relative to the inner disk to get the fraction of X-rays emitted into this latitude, $\cos\beta_x = 0.28$, yielding
	\begin{equation}
\begin{split}
\mbox{max}(L_{x,\mbox{\scriptsize eff}})
	& \lesssim 0.3\LEdd{} \times \cos\beta_x \times \dfrac{(\pi)(4.0 \Rsol{})^2}{4\pi a^2}\\
	& \lesssim 10^6 \Lsol{} \times 0.28 \times 0.02 \approx 100 \Lsol{}\\
\end{split}
\end{equation}
The companion's effective stellar luminosity falls within $\sim$500\,--1000$\Lsol{}$ based on the reported bolometric stellar luminosity 800\,--1600$\Lsol{}$ \citep{Soria_etal2001}.  More carefully integrating the incident-to-intrinsic ratio over the irradiated face (again, with gravity darkening) agrees closely with this simple estimate as the projected area and fraction of disk flux fall concurrently with (and faster than) the effective gravity toward L1.

	For irradiation operating alone, choosing the maximum observed luminosity as the reference point in eqn.\ref{eqn:MtotMod} would permit drops to $\approx$50\% of the observed maximum and only if the X-ray source were turned off completely, but this estimate is still fairly sensitive to companion temperature.  Harder and more isotropic X-ray flux from a hot corona may enhance modulation, but for LMC X-3 the maximum observed power-law flux is barely a fifth that of the disk, roughly equal to the projection factor reducing inner-disk flux onto the companion.  However, even if much deeper drops are possible, and irradiation-driven warping or some other mechanism were included to prevent the system from settling into a permanent steady high-soft state, the fact that SRM affects the flow at the very boundary means that any changes it introduces will suffer severe viscous dampening (\S\ref{sec:Res_noEC}).  Altogether, this suggests that SRM is significant, but certainly cannot explain the steep $\innrMddot{}$ declines by itself.
	
	Furthermore, we consider this simple model's predictions of the SRM magnitude an upper bound in light of as detailed two-dimensional hydrodynamic simulations of the envelope by \cite{ViHa_2007_2dSRM}.  They find that irradiation will still drive gas toward the nozzle, but the gas will also have ample time to cool down as it crosses the disk's shadow.  They note that because they do not solve for perpendicular velocity it may exceed their estimates near the nozzle, and we remark that warping of the outer disk might reveal more of the companion's equator and nozzle and negate the effects of cooling.  For our disk/corona simulations, we ignored the delay between irradiation and changes in $\Msdot{}$ since we estimated the sound-crossing time of the envelope near L1 to be $\approx$16 hr, far less than the viscous timescale.  However, in the case \cite{ViHa_2007_2dSRM} studied they found that some of the gas may take longer, up to several system orbital periods, to reach the nozzle.
		
		%\FloatBarrier
		\subsection{Bifurcation (DCB) and warping estimates}
		\label{sec:mechs_DCB}
		
	Matter streaming from the L1 point typically collides with the edge of the disk, which usually sits outside the circularization radius due to viscous spreading.  Because the disk is relatively cold at this radius, the collision is highly ballistic \citep{ArmLiv98}.  The fraction of matter streaming around the disk instead of immediately joining it can then be estimated simply by finding the altitudes at which the vertical disk and stream (both roughly Gaussian) density profiles match, and supposing \citep{Hessman99} that all the stream within this range immediately joins the disk while matter outside may stream further in.  This yields a streaming fraction,
		
\begin{equation}
f_s(t) = \mbox{erfc}\left[\left( \dfrac{\ln(\rho_{d0}/\rho_{s0})}{1-(H_s/H_d)^2} \right)^{1/2}\right]
\end{equation}
where $\rho_{d0}$ and $\rho_{s0}$ are disk and stream densities at $z=0$ and the stream scale height $H_s$ will not differ much from $\HsubL{}$---we also refer to Hessman (\citeyear{Hessman99}) for fits to the results of \citeauthor{LubShu75} (\citeyear{LubShu75}).  %As with irradiation, the physics of the immediate modulation is fairly simple, and the complication arises largely from the many dependences on the state of the companion and outer disk.

	Irradiation-driven warping of the outer disk may also affect the streaming fraction.  Again, \cite{OgDu01} and \cite{Foulkes_etal_2010_IrradWarping} suggest warping is possible in LMC X-3, and the latter work specifically finds a disk tilt of $10^{\rm o}$ likely for LMC X-3.  However, both use an isotropic central luminosity, and the latter use an Eddington ratio in luminosity for LMC X-3 comparable to our derived maximum Eddington ratio in $\innrMddot{}$, so we consider their results an upper bound on warping.

	We generalize the $f_s(t)$ estimate to a stream that scans the edge of a disk tilted by an angle $\vartheta_d(t)$ above the orbital plane.  Here, the vertical density centroid follows $z_0=R_d \sin(\vartheta_d(t) \cos(\Omega_{\mbox{\scriptsize syn}} t))$ where $R_d$ is the radius of the disk edge, and $\Omega_{\mbox{\scriptsize syn}}=\Omega_{K}(R_d)-\Omega_{\mbox{\scriptsize sys}}$ is the beat frequency between the Keplerian frequency at the disk edge and the system orbital frequency.  The finite travel time of the stream should add a roughly constant delay of order the local free-fall time, and for now we ignore this effect.  Assuming $d\vartheta_d/dt \ll \Omega_{\mbox{\scriptsize syn}}$, the altitudes of equal density are
\begin{equation}
\dfrac{z_{\pm}}{H_s} = \dfrac{z_0 H_s \pm H_d\sqrt{z_0^2+(H_s^2-H_d^2)\ln(\rho_{d0}/\rho_{s0})}}{H_s^2-H_d^2}.
\end{equation}
We will see that EC can depend very non-linearly on $f_s(t)$ at the boundary, but for now we use the orbit-averaged $f_s(t)$ as a gauge of plausible DCB strength:
\begin{equation}
\langle f_s(t) \rangle = \dfrac{1}{2}\left\langle 1+\mbox{erf}\left[\dfrac{z_-}{H_s}\right]+\mbox{erfc}\left[\dfrac{z_+}{H_s}\right] \right\rangle_{\phi}
\end{equation}
	
\begin{figure}[h]
	\centering
	\includegraphics[width=.5\textwidth]{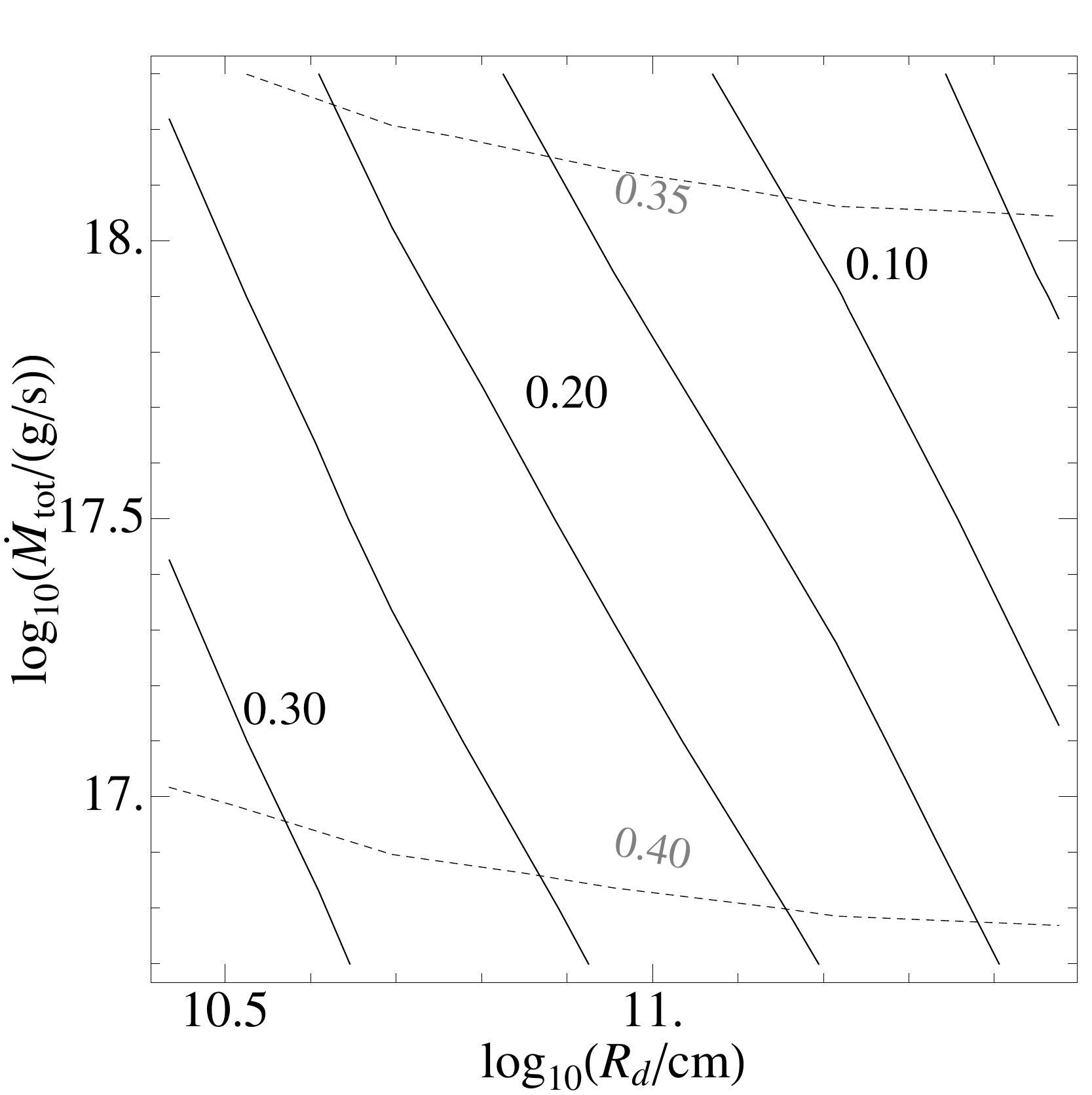}
	\caption{Solid and dashed contours show $\left\langle f_s\right\rangle$ for a disk edge tilted by $0^{\rm o}$ and $10^{\rm o}$ respectively, for a gas temperature of $16500$K, and range of relevant $\Mtot{}$ and outer disk radius.  Tilting the disk edge generally increases $\left\langle f_s\right\rangle$, but can also substantially change its dependence on the parameters, with the greatest effects at large $R_d$ and $\Mtot{}$.}
	\label{fig:fsContours}
	\end{figure}

	We plot $\left\langle f_s\right\rangle$ at the outer boundary for relevant ranges of total supply rate, $\Mtot{}$, and $R_d$, and for $\vartheta_d$ of $0^{\rm o}$ and $10^{\rm o}$ in fig.\ref{fig:fsContours}.  For an untilted disk, the contours are explained by the drop in disk scale height with radius and much slower drop with accretion rate, while for a tilted disk, the scanning greatly washes out the $R_d$ dependence leaving accretion rate as the dominant factor.  Our simple estimate also does not resolve the fate of the surviving stream beyond the edge (\citeauthor{Foulkes_etal_2010_IrradWarping} \citeyear{Foulkes_etal_2010_IrradWarping} do, but unfortunately not for LMC X-3 in particular), but should bound the fraction of mass diverted.
	
		%\FloatBarrier
		\subsection{CHW Prescription and Estimates}
		\label{sec:mechs_CHW}
		
	In Begelman et al. (\citeyear{BeMckSh83}), the authors considered an optically thin corona subject to Compton heating/cooling (ignoring bremsstrahlung and other heating/cooling mechanisms) and pointed out that accretion X-rays can heat the corona at all radii up to a temperature, $T_{\mbox{\scriptsize iC}}$, at which inverse-Compton heating and cooling equilibrate.  Whether a wind is launched at a given radius then depends mostly on whether this $T_{\mbox{\scriptsize iC}}$ is greater or smaller than the local virial temperature, $T_{\mbox{\scriptsize vir}}$, and the authors define a radius $\Ric{}$ by where the temperatures are equal, as well as a critical luminosity, $\Lcrit{} \approx \LEdd{}/33$ at which the gas can be Compton-heated to the virial temperature within the sound-crossing time of the local corona's scale height.  Because the tidal gravitational field falls off faster than the source luminosity, gas flows out most easily at large radii.  Though primarily a function of source X-ray luminosity and radius, the shape of the source spectrum can affect the mass-loss rate slightly but we ignore this effect.  \cite{BeMckSh83} computed mass-loss rates for total $\Mwdot{} \lesssim \Mtot{}$, while later work addresses dynamics and wind limit cycles \citep{ShMckLiBe_1986_CHW2}.
	
\begin{figure}
\centering
\includegraphics[width=.45\textwidth]{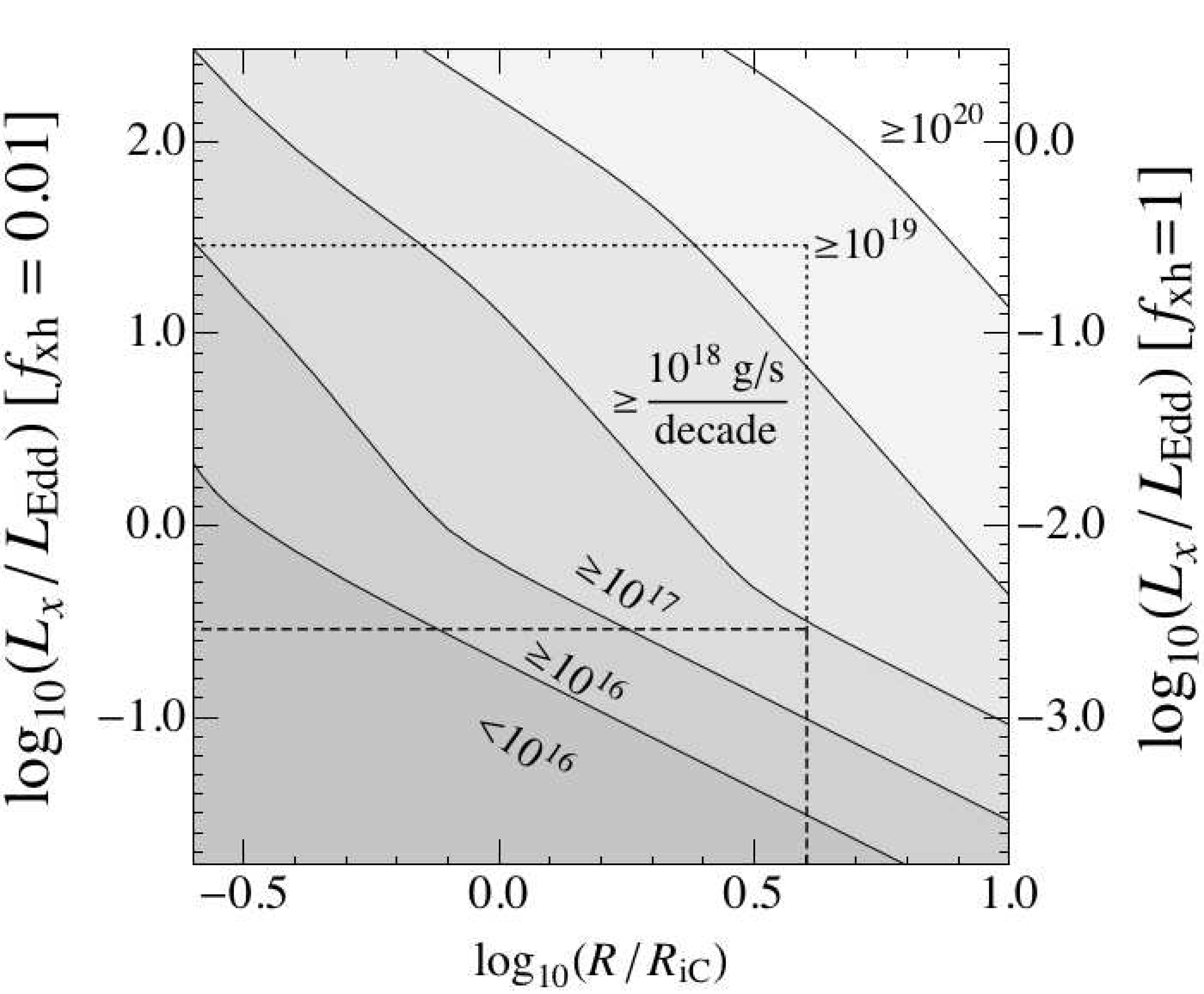}
\caption{Prescription $\Mdot{}_w$ losses per decade in $R/\Ric{}$ with left (right) vertical axis showing Eddington ratio when irradiation efficiency $f_{xh}$ is 0.01 (1).  The dashed (dotted) lines show a radial extent of $2\Rcirc{}$ and luminosity range for LMC X-3 with (without) reduced $f_{xh}$.}
\label{fig:Outflow}
\end{figure}

	\citeauthor{WoodsEtAl96} (\citeyear{WoodsEtAl96}) performed simulations to test the previous analytical prescription and amend it slightly---mostly by noting a shift in the location of $\Ric{}$ and providing corrections for low luminosities that do not immediately concern us.  We take their fitting formula for wind losses per unit area
\begin{equation}
\begin{split}
\dfrac{d\Mwdot{}}{dA} = &
	\mdot{}_{\mbox{\scriptsize ch}}\eta^{2/3}\left(\dfrac{1+(0.125\eta + 0.00382)^2/\xi^2}{1+\left(\eta^4(1+262 \xi^2)\right)^{-2}}\right)^{1/6}\\
&\times\exp\left[-(1-(1+0.25\xi^{-2})^{-1/2})^{2}/2\xi\right]\\
\end{split}
\label{eqn:WindRX}
\end{equation}	
	where the normalization $\mdot{}_{\mbox{\scriptsize ch}}$ is the ratio of corona pressure to sound speed at $\Ric{}$, $\xi = R/\Ric{}\approx 2 R/\Rcirc{}$, and their $\eta = L/\Lcrit{}$.  We then also introduce a factor $f_{xh}$ in $\eta\equiv f_{xh}L/\Lcrit{}$ for how well X-ray luminosity from an inner disk Compton heats the outer corona compared to the point source considered in the references.  Although the outflow geometry may permit parts of the outflow to eventually reach low inclinations relative to the inner disk, the chief hurdle is heating the gas when it is sitting deepest in the tidal gravity field.  Integrating $\cos i$ over the solid angle subtended by the outer corona versus half the disk's sky gives $f_{xh} \approx 0.025$.  This factor suppresses CHW considerably, while $f_{xh} \approx 1$ implies CHW will have significant effects at the maximum observed luminosities (fig.\ref{fig:Outflow}).  Furthermore, the $f_{xh}$ for depleting disk flow is likely different and smaller than the corona as the X-rays will have to reach higher inclination, and heat conduction from a transition layer will be competing with advection by the wind.  
	
	Winds may also be driven by other means, i.e., magneto-centrifugal and line driving, but extensive simulations by \citeauthor{Proga_2003_MHDwinds} \citeyearpar{Proga_2003_MHDwinds} with parameters relevant to LMC X-3 indicate that these losses in LMC X-3 will be at most a few percent of the total accretion rate.

	To gauge CHW self-screening, or screening the companion, consider a wind carrying away $10^{19}$g s$^{-1}$ (total, half this per disk face) at the local sound speed at $\Ric{}$.  If the density did not fall off with radius, the Thomson optical depth would be:
\begin{equation}
n \sigT{} \Delta s \approx \dfrac{0.5 \times 10^{19}\mbox{[g s$^{-1}$]}/m_p}{\Ric{2} \left(10^8 \mbox{[K]}\kboltz{}/m_p \right)^{1/2}} \sigT{} (a-\Ric{}) \lesssim 0.15,
\end{equation}
where $a$ is orbital separation.  That this extremely generous upper bound gives marginal absorption indicates the Compton wind will not screen the companion.  Instead, CHW and SRM will likely dampen each other's contribution to $\Mtot{}$-variability seen at inner radii as additional X-ray luminosity simultaneously increases $\Msdot{}$ supplied by the companion and $\Mwdot{}$ lost to space.  However, their interaction could enhance the scaling of $\Mddot{}/\Mcdot{}$ with $L_x$ at large radii.

		\subsection{PHII limits and discussion}
		\label{sec:mechs_PHII}
		
	The PHII is fundamental to the picture of transient BHXRBs and thus to future extension of our work, but the physics itself is not trivial to implement let alone fully understood as the (60 page) review by Lasota \citeyearpar{Lasota_2001_IoniInstRev} attests.  However, for LMC X-3, the strong, persistent disk emission should generally stabilize the disk within at least $1\Rcirc{}$, and we will later show (\S \ref{sec:Res_noEC}) that even drastic disk variability outside $\Rcirc{}/25$ is still too viscously dampened to explain observations, though the PHII may still contribute to the magnitude of disk variability, and likely plays an important role during complete state transitions.
	
	Taking either the mean or median of $\innrMddot{}(t)$ over our data set as a suitable proxy for supply rate gives $\langle \Msdot{} \rangle \approx 0.1 \MdotE{}$, and while the disk beyond $\sim\Rcirc{}/3$ will be cool enough to experience the PHII absent irradiation at $0.1\MdotE{}$ \citep[e.g. fig.1 of][]{JaniukCzerny_2011}, irradiation can stabilize more and possibly all of the outer disk \citep{Coriat_etal_12_PersistentVTransients}.  Work by \cite{Dubus_etal_1999_irrPHIIa} indicates that LMC X-3's disk would become susceptible to instability just beyond $\Rcirc{}$ at $10^{18}$g s$^{-1}$ for typical values of $\alpha$ (0.1) and an overall accretion to irradiation efficiency factor $\mathcal{C}$, originally fit to light curves of the BHXRB A0620-00 and roughly consistent with simple calculations based on an annulus-to-annulus irradiation geometry (see discussion in \citeauthor{Kim_etal_1999_TimeDepPHII} \citeyear{Kim_etal_1999_TimeDepPHII} and comparison at the end of \citeauthor{Dubus_etal_1999_irrPHIIa} \citeyear{Dubus_etal_1999_irrPHIIa} to \citeauthor{King_etal_1997_quickIrrLimits} \citeyear{King_etal_1997_quickIrrLimits}).  
	
	Because we do not see $\innrMddot{}(t)$ decay on $\Rcirc{}$-viscous timescales in LMC X-3, it appears that the PHII would also lack a large span of starved inner disk for a heating front to propagate through.  After the long, complete state transition of fig.\ref{fig:MdotsData} however, the disk recovery is flare-like, consistent with the notion that the PHII can play a significant role in LMC X-3 at low enough disk blackbody flux.
	
	%There are further wrinkles to the problem; \cite{Mescheryakov_etal_2011_irrRadTransport} show how scattering of X-rays in the disk atmosphere (let alone a corona), including harder X-rays that can more directly the inner disk, may significantly enhance stability while the changing height of the disk with radius (included in \cite{Dubus_etal_1999_irrPHIIa}) and warping can shadow or reveal disk material.

		\subsection{EC Background and Implementation}
		\label{sec:mechs_EC}

	As stated in \S\ref{sec:Intro}, EC may occur if the disk and corona are thermally coupled---if the disk cannot efficiently radiate away corona heat conducted onto it, nor sufficiently cool the corona via inverse-Compton cooling, then it will experience net heating and evaporate, but otherwise it cools the corona which then condenses onto it.  Thus the mass-exchange, or ``EC'' rate $\Mzdot{}$, is sensitively dependent on the balance of heating and cooling, and the very different scalings of heating and cooling mechanisms involved make possible a wide variety of behaviors.  At present, several EC models incorporate viscous and compressive heating, bremsstrahlung, and inverse-Compton cooling in the accretion flow including LTMHM07 and MP07.
	
	Besides separating the thresholds for disk formation/destruction normally degenerate under a bremsstrahlung-only density criterion via inverse-Compton cooling, and thus engender hysteresis \citep{Me_HoLiMe09}, it is also possible to evaporate the outer disk but condense it back onto the inner disk rapidly enough to drive correlated rises (and falls) of $\innrMddot{}$ with $\innrMcdot{}$ (again, fig.8 of MP07 and prominently in the left panel of our fig.\ref{fig:results_DfaultDeltaR}).  It is also possible to preferentially evaporate the middle of a disk to the point of destroying it as visible in \cite{MeLi_MH_07_RecondAndGap}, MP07, and several of our simulations.
	
	%Thus, accurately resolving the balance of heating and cooling in disk and corona is crucial in determining the mass-exchange, or ``EC'' rate $\Mzdot{}$.  At present, several EC models include viscous and compression heating, and bremsstrahlung and inverse-Compton cooling in the accretion flow to determine the resulting thermal conduction flux and the counteracting mass flux (e.g. LTMHM07, MP07).  The different scalings of these heating/cooling mechanisms and the strong dependence on current disk conditions make a large variety of phenomena possible.  We have already mentioned that condensation by inverse-Compton cooling from an extant disk separates the thresholds for disk formation/destruction that are usually degenerate under a bremsstrahlung-only density criterion, thus driving hysteresis \citep{Me_HoLiMe09}.  It is also possible to evaporate the outer disk but condense it back onto the inner disk rapidly enough to trigger sympathetic increases of $\innrMddot{}$ with $\innrMcdot{}$ (again, fig.8 of MP07 and prominently in the left panel of our fig.\ref{fig:results_DfaultDeltaR}).  This can be done to the point of destroying the middle portion of the disk as visible in \cite{MeLi_MH_07_RecondAndGap}, MP07, and several of our simulations.

	For our initial EC implementation, we do the following.  We assume azimuthal symmetry for the accretion flow and divide it into 45 logarithmically-spaced radial zones with a single virtual corona zone associated with each disk zone (i.e. the code is 1.5D).  We evolve the disk by solving mass fluxes with the standard viscous-disk equations \citep{FKR_book02} and a simple donor-cell scheme.  Meanwhile we assume that the local corona properties and EC rates match those of the steady-state corona and $\Mzdot{}$ solutions from LTMHM07 for the same local accretion rate and evolve the corona working inward from the outer boundary condition.  
	
	More specifically, for each disk zone we compute zone-boundary ($j\pm1/2$) velocities 
	\begin{equation}
	(v_d)_{j+\frac{1}{2}} = \dfrac{3 \left( (\nu \Sigma R^2 \Omega_K)_{j}-(\nu \Sigma R^2 \Omega_K)_{j+1} \right)}{2 R^2 \Omega_K \Delta R}
	\end{equation}
	with a viscosity based on a standard thermal equilibrium thin disk solution assuming Kramer's opacity as in \cite{FKR_book02} :
	\begin{equation}
	\nu_d = 2.13 \times 10^9 \left(\dfrac{\Mbh{}}{\Msol{}}\right)^{5/7} \alpha_d^{8/7} \left(\dfrac{3R}{R_S}\right)^{15/14} \Sigma_d^{3/7}.
	\end{equation}
	
	The overall disk and corona evolution is then governed by: 	
	\begin{equation}
	\Delta (\Sigma_d \Delta A)_j/\Delta t = (\Mddot{})_{j-\frac{1}{2}} - (\Mddot{})_{j+\frac{1}{2}} - (\Mzdot{\mbox{\scriptsize Rx}})_{j}
	\end{equation}
	and
	\begin{equation}
	(\Mcdot{})_{j-\frac{1}{2}} = (\Mcdot{})_{j+\frac{1}{2}} + (\Mzdot{\mbox{\scriptsize Rx}})_{j}
	\end{equation}
	
	where $\Delta A_j$ is the zone area.  The fluxes are also limited so as not to draw mass from a disk zone or the corona flow than is physically available, and the ``Rx'' emphasizes that we are plugging in the EC rates of LTMHM07 as a function of radius, and local $\Mcdot{}$ and effective disk temperature.  If $\Mcdot{}$ and $\Mzdot{}$ are anywhere comparable (of the same order of magnitude) to the viscous disk fluxes, then the computations are performed only once.  Otherwise, the latter two steps are relaxed further, and allowed to change $\Sigma_d$ but not $\Mddot{}$, until their iterations converge within a given tolerance (or exceed an iteration limit).  We note that large $|\Mzdot{}|$ can lead to oscillatory behavior with this simple scheme, which can be subdued but not fundamentally fixed with smaller time steps and tolerances.  This can be seen in the results of our simulations (fig. \ref{fig:results_DfaultDeltaR}-\ref{fig:CondVsHystComb}) as small sudden jumps in $\Mcdot{}$ (and somewhat in $\Mddot{}$ when EC is strong at small radii), but by changing time step and tolerance we have found that this does not impact the general, longer timescale features that immediately concern us.  We also explored small changes in the number of radial zones, obtaining similar results with 40 or more grid points, but our results diverged very quickly for coarser grids.
	
\begin{figure}[!ht]
\includegraphics[width=.45\textwidth]{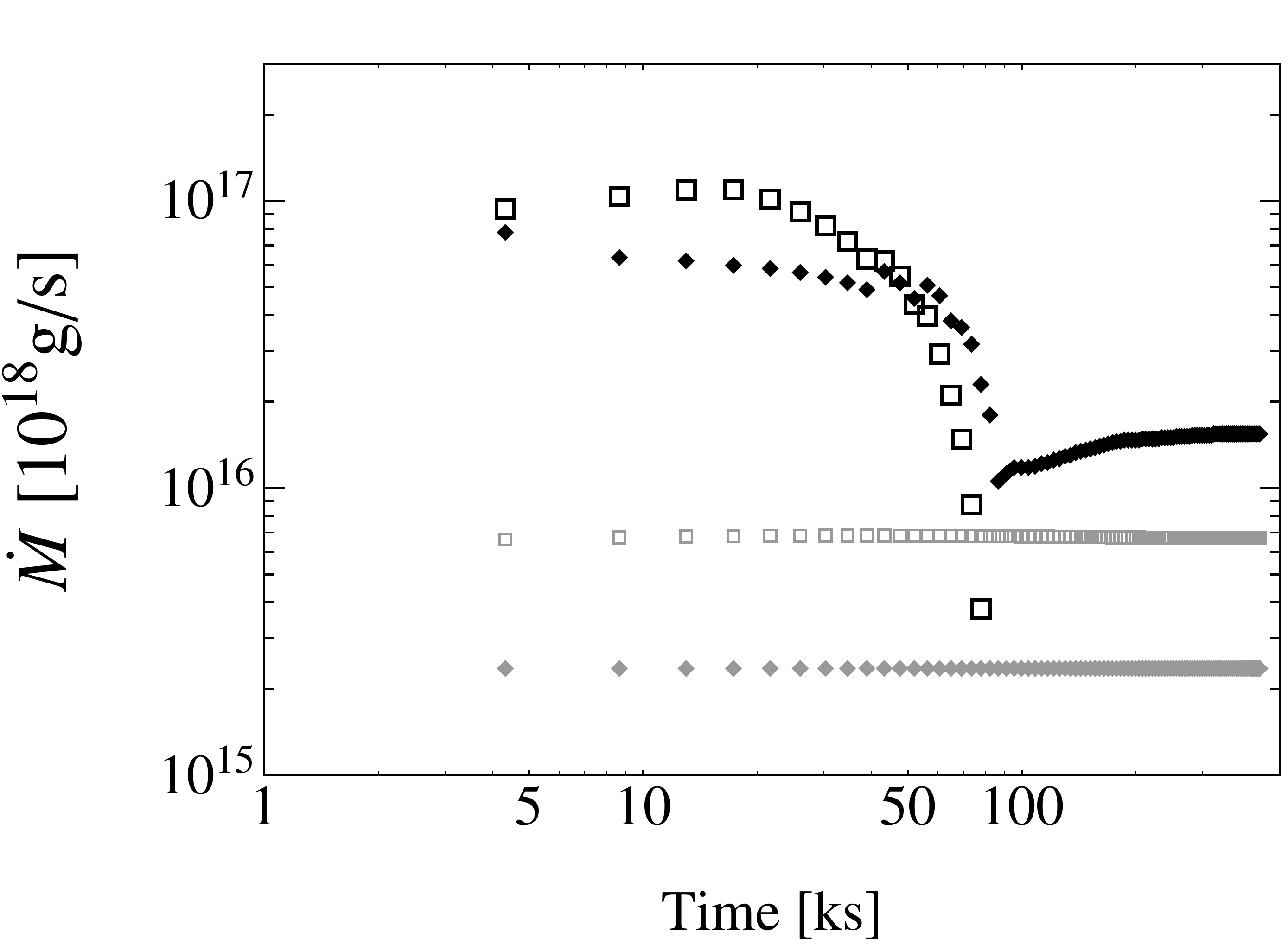}
\caption{The larger, darker empty squares and filled diamonds show $\innrMddot{}$ and $\innrMcdot{}$ respectively at the inner boundary for a run with modified heat conduction fluxes (see text \S\ref{sec:mechs_EC}), and the smaller, lighter symbols show the corresponding $\innrMddot{}$ and $\innrMcdot{}$ for default parameters - here we plot time logarithmically for more direct comparison with figure 8 of MP07.}
\label{fig:MP07_Comparison}
\end{figure}

	For steady input conditions, we confirm that our method does well reproducing cases considered by LTMHM07 (treating the disk fully lets it spread viscously to larger sizes mildly enhancing condensation).  To test our method's treatment of dynamic behavior, we looked at the same case MP07 studied: a disk spanning $3000$ Schwarzschild radii around a $10\Msol{}$ black hole with fixed boundary corona fraction $f_s$ of $0.1$ and mass supply rate of $10^{-3}$ times Eddington $\MdotE{}$.  Their time-dependent method evolves the corona self-consistently on its dynamical timescale making it more physically realistic, but this also requires many more time steps.  We found that with default parameters and physics our code never evaporates any part of the disk, while MP07 predict the formation of a gap that eats its way inward.  However, we also found that if we scaled up the heat-conduction fluxes predicted by LTMHM07 for zones where inverse Compton cooling does and does not set the electron temperature by a factor of three and five respectively (adjusting the formula identifying the zones accordingly) then we do obtain good agreement with MP07 (see fig.\ref{fig:MP07_Comparison}, and their fig.8).  By preferentially scaling up heat fluxes in the zones where inverse-Compton cooling limits electron temperature we obtained more rapid evaporation starting further inward while doing the same for heat fluxes in non-Compton zones led to increased stability.  In this particular case, we saw little change when lowering the magnetic-to-gas pressure ratio, $\beta_c$, a global constant in LTMHM07, from 0.8 to 0.1.
	
\begin{figure*}[!ht]
	\centering
	\includegraphics[width=0.60\textwidth]{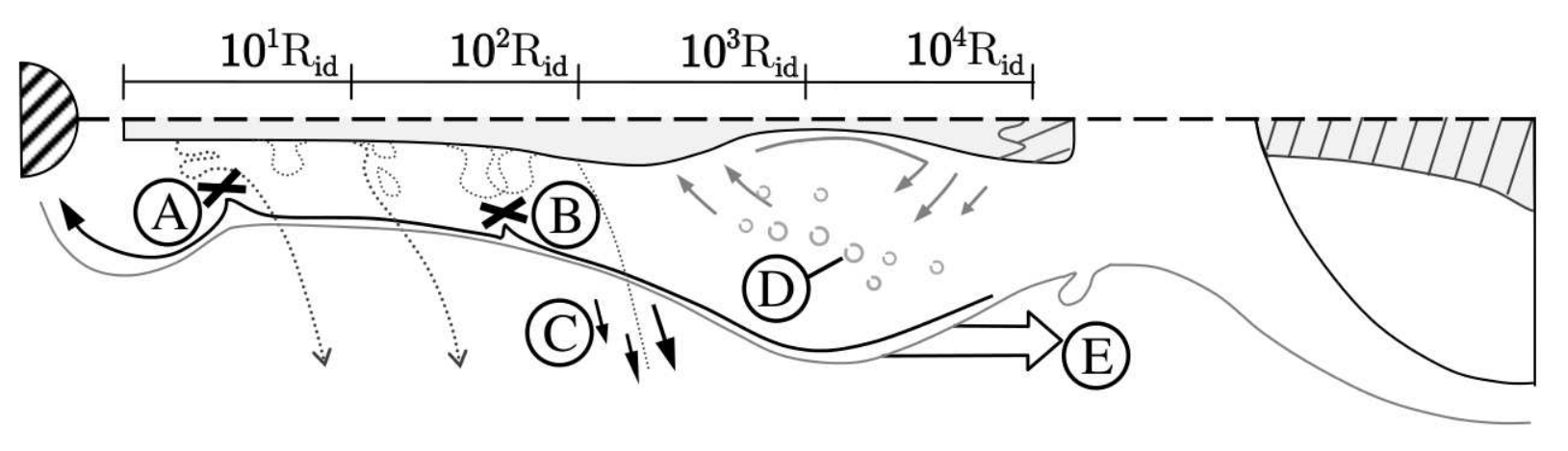}
	\caption{A cartoon summarizing possible complications to the model especially concerning EC (\S\ref{sec:mechs_EC},\S\ref{sec:DiscConc}).  At small radii, the magnetic field (dotted lines) in the corona and at the disk-corona interface may be significantly non-poloidal thus suppressing condensation (A), and further outward buoyant magnetic loops may still alter the magnetic field geometry besides introducing additional reconnection heating to the corona to continue supressing condensation (B).  MHD winds might also be stronger than predicted and carry away more of the corona (C), while clumping of the corona at large radii may instead enhance condensation over evaporation there (D).  The potential for the corona to viscously outflow radially wherever it achieves large density gradients (E) may also significantly affect our results.}
	\label{fig:MechComplications}
	\end{figure*}

	Both the models of LTMHM07 and MP07 necessarily neglect, or precede, some additional physical effects which may be relevant to our early results so we discuss them here (and summarize in fig.\ref{fig:MechComplications}) to motivate our more exploratory simulations (\S\ref{sec:Res_EC} and fig.\ref{fig:CondVsHystComb}).  In both models, condensation is a smooth, unresolved flow, but applying the results of \citet{WaChLi_2012_ClumpyAccr} shows that the corona is liable to clump at radii greater than roughly 100$R_{id}$ under typical conditions for LMC X-3.  Such clumping potentially increases cooling (thus condensation) efficiency.  Both LTMHM07 and MP07 use Spitzer electron conduction throughout the problem domain, and both suspect that the effective thermal conduction coefficient $\kappa$ may be significantly smaller.  Although the degree of tangling in the magnetic fields of the transition zone is far harder to constrain, it is amenable to parameterization.  Meanwhile, \cite{Cao_2011_AdafBfields} provides a recent calculation for how much the ordered component of field shifts from predominantly poloidal outside $\sim 10 R_{id}$ to predominantly toroidal inside.  Lastly, neither model includes mechanisms to spontaneously produce corona, a point MP07 especially emphasize.  Indeed, since \cite{Galeev_etal_79_diskBfields} derived that within a certain radius, the buoyancy of magnetic loops formed within the disk can outpace their reconnection leading to a carpet of buoyant loops, this solar-like corona has often been invoked as a partial or complete source of corona.  For LMC X-3, the condition on radius in \cite{Galeev_etal_79_diskBfields} gives $R\lesssim 300(\MdotE{}/\innrMddot{})R_S$.  It is hard to imagine this mechanism alone generating $\innrMcdot{}$ pulses that anticipate $\innrMddot{}$ drops lasting substantially longer than the viscous timescale at $\sim 100 R_S$, but it may play an important role by reheating and replenishing the corona, and affecting the local magnetic field geometry.
	
	An important caveat in applying the LTMHM07 model came to our attention after running our simulations.  For $\Mcdot{}\gtrsim$ 0.1 Eddington near $R\sim$100$R_S$, the conduction or Compton-cooling (at high $\innrMddot{}$) limiting temperatures may cross the coupling temperature (e.g. \citeauthor{BradFrank2009} \citeyear{BradFrank2009}).  Under these conditions, the model will tend to overpredict condensation and thus exaggerate the amplitude and duration of the sympathetic mode, but triggering the effect requires strong, correlated $\innrMcdot{}$, $\innrMddot{}$ to already be underway.

	\section{Simulation results}
	\label{sec:Rez}
	
		\subsection{Models without EC}
		\label{sec:Res_noEC}
		
	To illustrate how a combination of non-EC mechanisms has difficulty explaining the magnitude of $\Mddot{}$ drops and especially the pattern of rapid decline versus slow recovery, we employ a very simple model where disk accretion is computed using the analytical machinery of \cite{Zdz_etal_2009_visc}.  The latter is derived assuming a time-independent viscosity with power-law dependence on radius, and we referred to the thin-disk solutions on pg.93 of \cite{FKR_book02} for all relevant viscosity parameters.  This method prevents incorporating radial dependence of \Mwdot{}, but because wind losses fall rapidly with decreasing radius, and since we predict they are fairly weak anyway, assuming that they take place near the boundary does not invalidate the main results of this toy model.  This method also precludes incorporating he PHII, but as discussed in \S\ref{sec:mechs_PHII} the main effects of the PHII should typically be limited to radii beyond $\sim\Rcirc{}$ in LMC X-3.
	
\begin{figure*}[!htb]
	\centering
	\includegraphics[width=1.\textwidth]{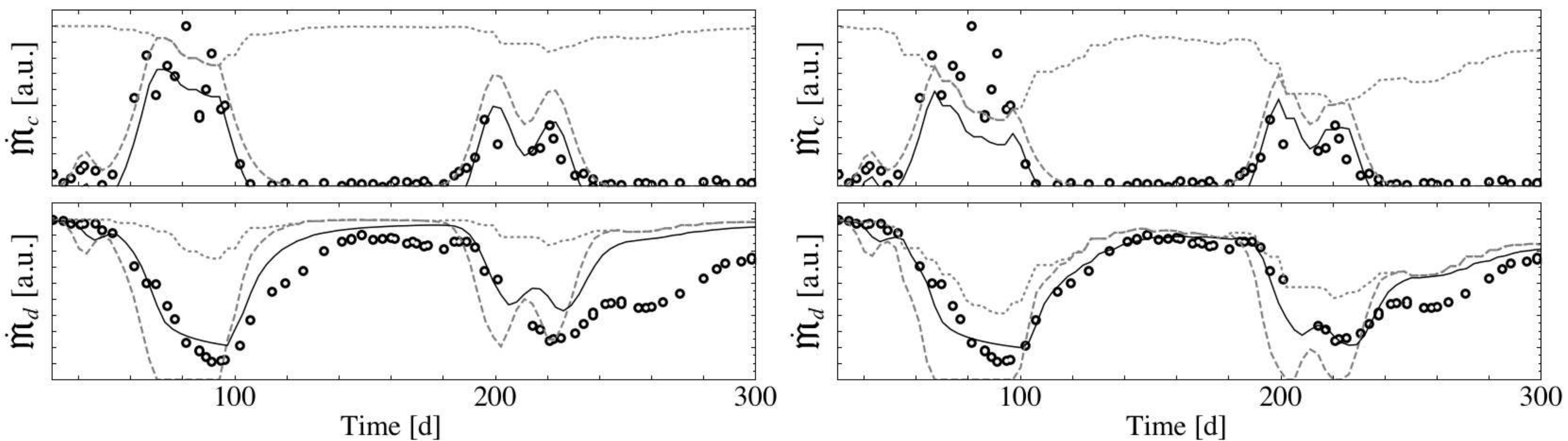}
	\caption{The solid curves in the bottom (top) panels show $\innrMddot{}$ ($\innrMcdot{}$) simulated with the simplified model discussed in \S\ref{sec:Res_noEC} and empty circles show the observed $\innrMddot{}$ ($\innrMcdot{}$), where the simulation units first are chosen to match the observed and simulated initial $\innrMddot{}$, as the simple model's disk machinery has no direct dependence on absolute accretion rate.  The data points in the top panels are then scaled so that the maximum observed $\innrMcdot{}$ equals the initial $\innrMddot{}$ which is a much larger absolute scale than our estimates suggest.  The right panels are for a run with greater variability in the total mass supply which is shown as a dotted curve in all panels.  The dashed curves show the effective corona/disk inputs at the outer boundary so that the remaining difference between solid and dashed curves in top panels indicates the wind loss.}
	\label{fig:nAwBwC}
	\end{figure*}

	The left panel in fig.\ref{fig:nAwBwC} shows a calculation with this reduced model geared toward reproducing the first disk drop in fig.\ref{fig:MdotsData}.  We set $f_s$ manually, and SRM and wind losses were also computed beforehand as functions of the observed $\innrMddot{}$.  To reproduce the depth of the first drop, we first allowed sustained $f_s$ of 100\% and when this proved insufficient we moved the outer disk radius inward to a mere 0.04$\Rcirc{}$ for the runs shown, still employing $f_s$ of 100\%, and leading to massive $\innrMcdot{}$ pulses compared to our estimates.  This can be corrected somewhat by invoking a wind stronger than our estimates.  

	Besides requiring unrealistic values with respect to our estimates, and an extreme ad-hoc disk truncation, the toy model resists efforts to simultaneously improve agreement with other major features of the data.  To improve model-data agreement for the second disk drop in the left panel of fig.\ref{fig:nAwBwC} without increasing disk radius (which would obviously undo agreement with the first $\innrMddot{}$ drop) requires either increased SRM or increasing the height and duration of the second coronal pulse.  The latter will generate obvious disagreement with the second $\innrMcdot{}$ pulse by attaching a tail that is very clearly not observed; the former significantly reduces the amplitude of the first coronal pulse relative to the second so that one must invoke a stronger and more complicated wind mechanism.
	
	Again, the chief problem is that a disk flow will be viscously smeared too much to match observations unless variability is driven at a relatively small radius where even maximally efficient CHW should be insubstantial, and the PHII should not operate nor regularly drive heating fronts.  In the next section, we will show how EC may introduce a evaporation-to-condensation transition or gap at radii comparable to the outer edge of the arbitrarily truncated disk of the toy model, but that it also tends to overpredict condensation at inner radii, generating correlated $\innrMddot{}$-$\innrMcdot{}$ rises and falls inconsistent with observations.

		\subsection{Models including EC}
		\label{sec:Res_EC}
	
	Except where specifically noted otherwise, for these simulations we again use $\alpha_d=0.1$, but a corona $\alpha_c=0.2$, the standard Spitzer coefficient for electron thermal conduction, a $\beta_c=0.8$ for the LTMHM07 EC prescription, the observationally favored $\Rcirc{}=2.7\times10^{11}$cm, assume $R_{id}$ is 3$R_S$, and set $f_{xh}=0.03$.  We note here that the SRM results from \S\ref{sec:mechs_SRM} specifically correspond to a value of $L_{x,\mbox{\scriptsize eff}}^{\mbox{\scriptsize ref}}/L_{\star,\mbox{\scriptsize eff}}^{\mbox{\scriptsize ref}}=0.4$ for $\innrMddot{}=10^{18}$g s$^{-1}$, and to $6.5$ scaleheights between the nozzle and the stellar surface for $T_{\star}^{\mbox{\scriptsize ref}}=16500$K.
	
\begin{figure*}[!htb]
	\centering
	\includegraphics[width=.9\textwidth]{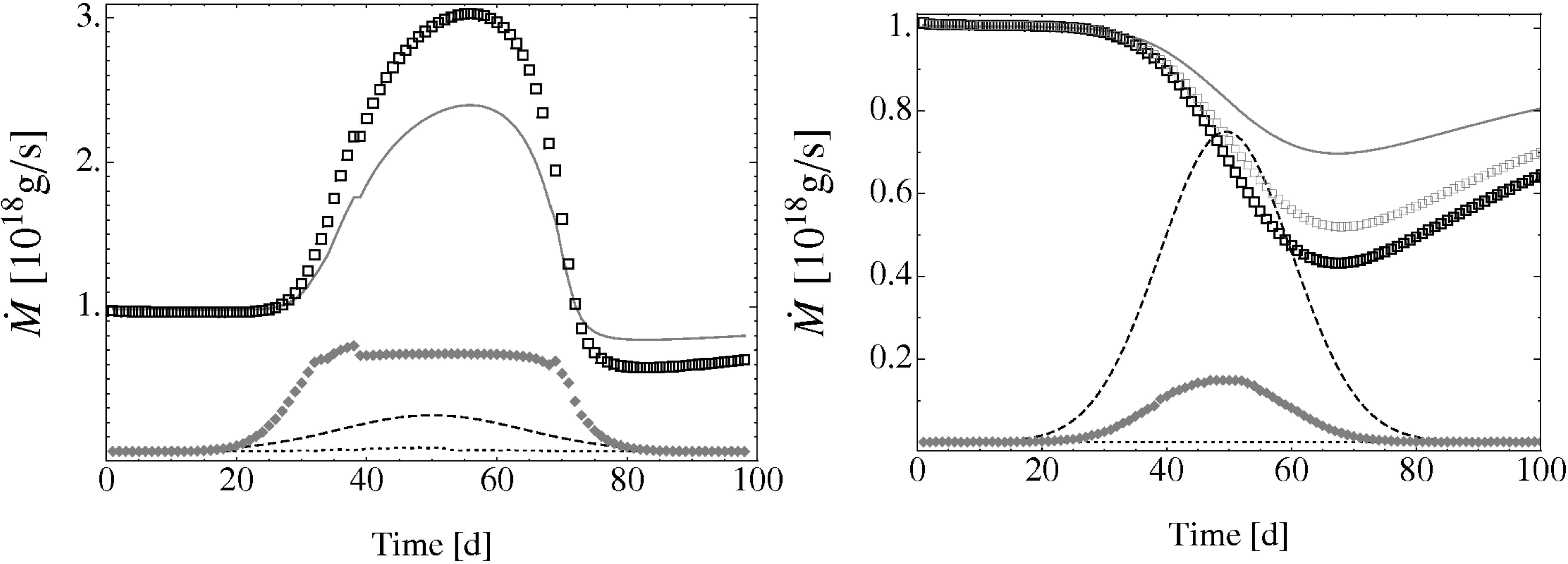}
	\caption{Results with rough EC-implementation showing $\innrMddot{}$ (empty black squares), $\innrMcdot{}$ (filled diamonds), $\Mtot{}$ (solid line), $\Mwdot{}$ (dotted line), $f_s\times 10^{18}$g s$^{-1}$ (dashed line) while $\innrMddot{}$ without EC turned on (gray empty squares) is included for the right panel.  The run on the left uses the full circularization radius and shows strong condensation while the right panel shows a run with a disk size of $3\times 10^{10}$cm. (\S\ref{sec:Res_EC}).}
	\label{fig:results_DfaultDeltaR}
\end{figure*}

	For the EC-inclusive model, we first simulated a disk with standard density profile extending to the circularization radius reacting to a mild Gaussian $f_s(t)$ pulse, and show the results in the left panel of fig.\ref{fig:results_DfaultDeltaR}.  Two immediately remarkable features include the sympathetic rise and fall of $\innrMddot{}$ with $\innrMcdot{}$, and the general saturation of $\innrMcdot{}$ response when the outer disk is most intensely siphoning onto the inner disk.  This saturation physically arises from the steep transition between evaporation and condensation.  In the mass exchange model of LTMHM07 that we employ, the local evaporation/condensation rate $\Mzdot{}$, varies with the local corona accretion rate in terms of Eddington ratio, $\mcdot{}$, like
\begin{equation}
\Mzdot{} \sim a\mcdot{7/5}(1-b\mcdot{20/21})
\label{eqn:ECpolynomial}
\end{equation} 
where $a$ and $b$ are functions of many other parameters, local variables, and radius itself.  If these other variables vary weakly with radius, then a very dense corona at some radius will lead to efficient condensation slightly further inward, and subsequent changes in $\Mzdot{}$ about zero will be driven by the weaker variations in the critical value of $\mcdot{}$.

%However, we must also point out that our code tends to exaggerate the severity of this saturation slightly, as well as the sharpness of jumps in $\innrMcdot{}$.  We recall that the corona is effectively an infinitely fast stream in our code, and our relaxation scheme has finite tolerance.  We confirmed that changes to the latter and the number of grid cells altered these small percent-level features, but not the significant features we discuss. 

	The issue of sympathetic accretion prompted us to consider a scenario in which the outer disk mass most vulnerable to being siphoned has already been evaporated away, such that there is a gap in the outer disk.  We first studied the effects of simply truncating the disk, and show an example with radius $3\times 10^{10}$cm and default LTMHM07 EC parameters in the right panel of fig.\ref{fig:results_DfaultDeltaR}.  Truncating the disk to this radius prevents triggering sympathetic accretion while generating appreciable $\innrMddot{}$ variability and diminishing, but not eliminating, EC's role in amplifying variability over the simulation domain.

\begin{figure*}[!htb]
	\centering
	\includegraphics[width=.9\textwidth]{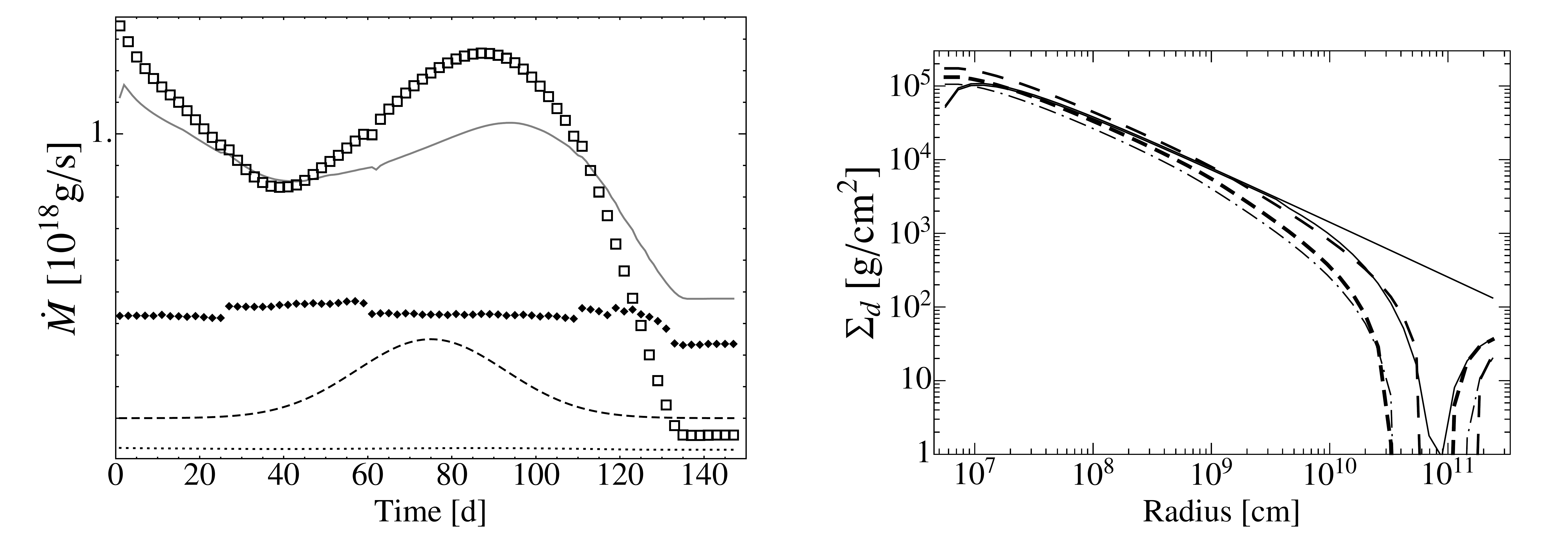}
	\caption{Accretion history (left, see fig.\ref{fig:results_DfaultDeltaR} caption for symbol meanings) and evolution of the disk surface density profile (right) with snapshots at 80, 100, and 120 days shown in long-dashed, thick-dashed, and dot-dashed lines while the thin solid curves show the gap initial conditions relative to the standard thin disk.}
	\label{fig:ETB3comb}
\end{figure*}

% no const.fig. version
	We next attempted to generate this gap self-consistently, while preserving the interior aspect of the accretion flow that works fairly well.  To this end, we first subjected a full disk to a constant $f_s$ at the boundary.  The run confirmed that a small seed corona can rapidly evaporate the outer disk, that this corona is immediately condensed only slight inward, but also that the process of forming a complete gap would take on the order of years in our standard EC implementation.  To pursue this idea further, we ran simulations with a gap already inserted of which fig.\ref{fig:ETB3comb} is representative.  Besides the initial relaxation of $\innrMddot{}$ due to viscous spreading of the inner disk exceeding and resupply via condensation, one can see that sympathetic accretion is still an issue, and $\Mcdot{}$ variability inward of the gap is severely suppressed again, as the high corona fraction of the gap triggers the saturation effect described above.

\begin{figure*}[!htb]
	\centering
	\includegraphics[width=.9\textwidth]{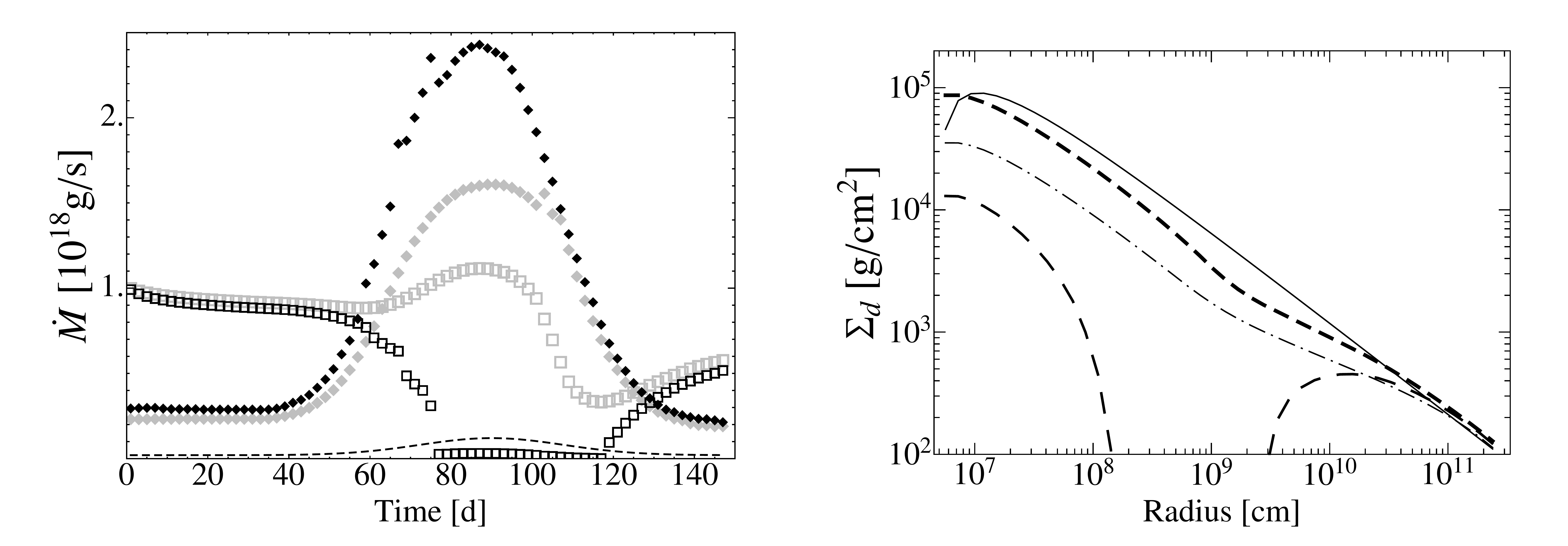}
	\caption{The left panel shows accretion histories for $\innrMddot{}$ (empty squares) and $\innrMcdot{}$ (diamonds) in cases A (thin, black), and B (thick, light gray) as described in \S\ref{sec:Res_EC} where $f_s\times 10^{18}$g s$^{-1}$ is also shown again, but $\Msdot{}$ and $\Mwdot{}$ are omitted.  The right panel shows evolution of disk surface density profile for case A at 60, 90, and 120 days with the same convention as in fig.\ref{fig:ETB3comb}.}
	\label{fig:CondVsHystComb}
\end{figure*}

	Since our standard model and implementation of EC faces fundamental problems in reproducing major features of the data, we studied the effects of introducing physically motivated, if not yet rigorously justified modifications.  Thus far, it appears that the most successful modifications follow a fairly strict pattern of effectively raising the coronal heating and critical evaporation-to-condensation $\mcdot{}$ over the innermost two decades in $R/R_{id}{}$.  The latter is nearly inversely proportional to $b$ in eqn.\ref{eqn:ECpolynomial} which in the model of LTMHM07 scales with the corona viscosity parameter $\alpha_c$, radius, electron thermal conduction coefficient $\kappa$, and gas-to-total pressure ratio $\beta_c$ as
\begin{equation}
b \sim \beta_c\kappa^{1/5}\alpha_c^{-14/15}\left(\dfrac{R}{R_{id}}\right)^{-1/10}
\label{eqn:EC_b_term}
\end{equation}
though we should point out that $\beta_c$ enters their model strictly via a prescription for compressive heating.  In the simulations producing fig.\ref{fig:CondVsHystComb}, we raised $\alpha_c$ linearly from 0.2 to 0.4 inwards over $100 R_{id}$, and independently adjusted $b$ over radial zones spanning $10^{0}$--$10^{2.7}$, $10^{2.7}$--$10^{3.4}$, and $10^{3.4}$--$10^{4.6}$ in $R/R_{id}$.  For run A in fig.\ref{fig:CondVsHystComb}, we scaled $b$ in these zones by 0.1, 0.2, and 0.4 respectively, and for run B by 0.4, 0.6, and 0.8.  Although the simulated $\Mcdot{}$ pulses are far more massive than observations indicate, these modifications very clearly control the degree of hysteresis versus sympathetic accretion.  If the overall magnitude of EC is smaller, then for higher $f_s$, and/or SRM moderately larger than expected, this modified model could reproduce the observed hysteresis.

	The most conceivably adjustable parameters in eqn.\ref{eqn:EC_b_term} are $\alpha_c$, especially if understood to include other heating mechanisms, and $\kappa$.  Since the current contrast between simulation and observation still favors weaker EC, the requirement on additional heating would not likely be as extreme as implied by our modifications, but this then requires even greater deviation in $\kappa$ which affects $b$ rather weakly, and these parameters are not necessarily independent in reality.
	
	\section{Discussion and conclusions}
	\label{sec:DiscConc}
	
	Thus far, we still cannot offer a very definitive solution for LMC X-3's behavior, but we have more rigorously examined several physical mechanisms popularly invoked to explain variability in LMC X-3 and additionally considered evaporation and condensation.  We have found that if condensation is suppressed at inner radii (or over-predicted in the current model), then EC may naturally reconcile the observational evidence for RLO accretion and associated circularization radius, as well as the large amplitude, long duration, and rapid declines in hysteresis episodes with our estimates of other major viscous timescale variability mechanisms and the viscous dampening that they would undergo.
		
	However, we wish to emphasize that our current EC implementation and the steady-state theory informing it by default led to excessive condensation when compared as accurately as possible to observations, and it led to significantly different predictions for the particular low-$\Msdot{}$ case studied by MP07.  As discussed in \S\ref{sec:mechs_EC}, both the LTMHM07 and MP07 models necessarily neglected some physics, some of which might enhance heating and/or suppress conduction closer to the black hole, and thus help explain the discrepancy between observations and the predictions of our code with the default EC prescription.
	
	To this end, we have reproduced and are testing a code that follows MP07 and evolves the disk and corona mass and energy equations self-consistently.  An additional advantage is that we can naturally include physics behind the PHII by incorporating detailed results for disk cooling as a function of density and central temperature from previous work - crucial to studying transient systems.  However, this explicit method suffers from advancing by a very small time step.  We have started building a parallelized implicit method which we hope to develop further during simulations with the explicit method, but we also hope to find ways to save on excessive computation through better physical understanding of the problem.

	Our estimates for the reduced efficiency of CHW, and cursory examination of theory results for MHD winds in similar systems.  \cite{Proga_2003_MHDwinds} suggest that winds in general will have little affect on accretion dynamics in LMC X-3.  However, we will continue to consider how efficiency of CHW might be increased, or that MHD winds may be stronger than anticipated \citep[e.g.][]{King_etal_2012_BWindsInJ17091}.

	Other assumptions in our current modeling that bear repeated mention include the alpha-prescription disk, and how we infer and interpret the disk and coronal accretion rates.  Regarding the former, it is at least expected that under conditions associated with jet flow, angular momentum transport via the magnetically driven outflow may become substantial or dominant compared to viscous transport (i.e. the magneto-rotational instability) at least within the inner flow \citep[e.g.][]{Zanni_etal_2007_specJetSims,CasseFerreira_2000_ColdWindCondsIV}.  Evolution of the magnetic fields in the disk may also lead to intrinsic variability out to $\sim$100$R_{id}$, as in \cite{deGuiranFerreira_2011_arXiv_Bdrag}.  The potential for corona flow to stall centrifugally as a function of external flow conditions and local viscosity \citep[e.g.,][]{ChakrabartiTitarchuk_1995_TwoComp,Garain_etal_2012_2FlowMonteCarlo1} might be realized in LMC X-3, but is also usually expected to occur well within the innermost 100$R_{id}$ of the flow and to destroy the disk interior to the centrifugal shock, so it may be most relevant to state transitions.  If more frequently prevalent though, the latter could alter our picture of spectral production, enhance $\innrMcdot{}$ variability especially on shorter timescales, and change the dynamics of EC.  Based on the pattern of the power-law component to anticipate declines in blackbody flux and the viscous recovery timescale of the blackbody component, we are still naturally inclined to favor a picture where variability is driven outside-in so that mechanisms like these and EC would predominantly accelerate and enhance $\innrMddot{}$ declines driven by known mechanisms operating in the outer flow.%  We disfavor, but do not rule out that mechanisms in the inner flow might drive modulation with a timescale much greater than the local viscous timescale.

%	Some other justified but pivotal assumptions in our modeling that deserve repeated mention include the alpha-prescription disk, and how we infer and interpret the disk and coronal accretion rates.  Regarding the former in particular, it is at least predicted that in the innermost 100$R_{id}$ of the flow under conditions associated with jet outflow, angular momentum transport by outflow will become substantial or dominant, and evolution of magnetic fields in the disk may intrinsically lead to variablity as in \cite{deGuiranFerreira_2011_arXiv_Bdrag}.  The potential for the corona to be stalled or ``shocked'' centrifugally as a function of external flow conditions \citep[e.g.][]{ChakrabartiTitarchuk_1995_TwoComp,Garain_etal_2012_2FlowMonteCarlo1} might be realized in LMC X-3, and alter our picture of spectral production besides influencing the dynamics of EC and enhancing corona variability relative to the disk.  %As mentioned in \S\ref{sec:mechs_EC} in connection with theories of coronal production, based on the pattern for the power-law component to anticipate the declines in blackbody flux and the viscous recovery timescale of the blackbody component we are naturally inclined to favor a picture with one or more fairly strong mechanisms operating at larger radii and associated viscous timescales versus more sensitive mechanisms further inwards.
	
	Our immediate focus will be resolving the outstanding questions regarding the current mechanisms considered, especially evaporation and condensation.  After understanding and constraining these better, we hope to extend our investigations to additional physics, systems, and phenomena, especially the transients and jet launching.

	\section{Acknowledgements}
	\label{sec:Ack}
	
The authors acknowledge support through the NASA ADP program grant NNX09AC86G.  The authors would also like to thank the referee for suggestions that improved the content and clarity of the paper.
	
\begin{appendix}	

	\section{Relating accretion rates to spectra production}
	\label{sec:Appendix_Norms}
	
	For the blackbody spectrum, \citet{Zimmerman_etal_2005_bbSpec} note that \texttt{XSPEC} fits the maximum temperature and normalization constant to a temperature profile of the form $T(R) = T_{\mbox{\scriptsize max}} (R_{id}/R)^{3/4}$.  The fit to the peak temperature using this profile is only $\sim 5\%$ smaller than the peak temperature found fitting a temperature profile based on a zero torque boundary condition and color-correction factor $\fcol{} $ \citep{EbHaSu84}:
\begin{equation}
\Tbb{}(R) = \fcol{} T_{\mbox{\scriptsize eff}}(R) = T_{*} (R_{id}/R)^{3/4}\left( 1-\left(\dfrac{R_{id}}{R}\right)^{1/2}\right)^{1/4},
\end{equation}
where
\begin{equation}
T_{*} = \fcol{} \left(\dfrac{3 G\Mbh{}\innrMddot{}}{8\pi\sigSB{} R_{id}^3}\right)^{1/4} = 2.05 T_{\mbox{\scriptsize max}}.
\end{equation}
Because there is little difference in the fitted peak temperatures, because we plan to fix the black hole mass and $R_{id}$, and because we prefer the physically-motivated temperature profile we will use it instead.  Integrating over disk annuli then gives the familiar formula for flux:
\begin{equation}
F^{\mbox{\scriptsize mbb}}_{\nu} = \dfrac{1}{\fcol{4}} \dfrac{4\pi h \cos i \cdot \nu^3}{c^2 \dsys{2}} \int_{R_{id}}^{R_d} \dfrac{R dR}{\exp(h\nu/\fcol{}\kboltz{} T_{\mbox{\scriptsize eff}}(R))-1}
\end{equation}
	
	Because \citet{ShiTak95} predict that $\fcol{}$ depends weakly on radius, accretion rate, and other parameters, we fix $\fcol{}=1.7$.  This then implies that the maximum disk accretion rate ranges from 0.07--0.29$\MdotE{}$ for $R_{id}$ spanning 1 to 3 Schwarzschild radii.  We scale the $\innrMddot{}$ of fig.\ref{fig:MdotsData} by matching the observation with the highest blackbody flux and temperature to the maximum 0.29$\MdotE{}$.  We note that if dissipation interior to the last stable circular orbit is significant, this will also put the actual $\innrMddot{}$ below our estimate \citep{BeHaKr_2008_RinnrRad,ShNaMC_2008_RinnrRadToo}.
	
	Inferring  $\innrMcdot{}(t)$ requires additional assumptions but many are well constrained within ADAF theory \citep{NaYi95b}.  Specifically, theory predicts that corona ions are very effectively virialized at inner radii and much hotter than the electrons, whose exact temperature depends on many conditions, but is generally flat over the innermost $100 R_{id}$ and of order $100$keV in the cases considered by \cite{NaYi95b}.  The former means that we can confidently predict scale height given corona $\alpha_c$ while the latter provides some justification for choosing a constant electron temperature in corona emission calculations.  Taking $R_{id}=3R_S$, corona alpha parameter $\alpha_c=0.2$, and gas-to-total pressure ratio $\beta_c=0.8$, we obtain \citep{NaYi95b} the following estimates for corona density $n_c$, corona height $H_c$, and coronal optical depth $\tau$ in terms of $\mdot{}_c=\innrMcdot{}/\MdotE{}$,

\begin{equation}
n_c(R_{id},\mdot{}_c) \approx 1.2\times 10^{19}\mdot{}_c(\Msol{}/\Mbh{})\mbox{[cm$^{-3}$]},
\end{equation}

\begin{equation}
H_c(R_{id},\mdot{}_c)/R_{id} = h_c \approx 1 ,
\end{equation}

\begin{equation}
\tau(R_{id},\mdot{}_c) \approx 58 \mdot{}_c
\end{equation}
However, we remind the reader that the latter is fairly sensitive to $\alpha_c$, scaling roughly like $\alpha_c^{-1}$.  If we take the scattering fraction to be $\mathcal{P_{\tau}} = 1-e^{-\tau}$ then based on the scattering fractions returned by XSPEC for the mixed states, inversion gives us the simple $\innrMcdot{}$ estimate shown as the solid line in fig.\ref{fig:MdotsData}.  

We compare this simple estimate with a more detailed power-law flux calculation.  \cite{HuaTitarchuk_1995_iC} find a Green's function for the output energy spectrum given the seed photon energy spectrum per scattered seed photon (specifically, their eqn.9), $G_{\nu}(x,x_s,T_e,\mathcal{P}_{\tau})$.  The latter depends again on scattering fraction, as well as corona electron temperature $T_e$, and the output ($x$) and seed ($x_s$) dimensionless photon energies ($x_*=h\nu_*/\kboltz{}T_e$).  Using the result of a more detailed calculation for the scattering fraction in a slab geometry from \cite{Zdz_etal_1994_iC_app},
\begin{equation}
\mathcal{P}_{\tau} = 1+\frac{1}{2}e^{-\tau}\left(\frac{1}{\tau}-1\right)-\frac{1}{2\tau}+\frac{\tau}{2}\mbox{Ei}(1,\tau),
\label{eqn:ZdzPtau}
\end{equation}
we convolve $G_{\nu}(x,x_s,T_e,\mathcal{P}_{\tau})$ with the flux of seed photons that can scatter (overall $\mathcal{P}_{\tau}$ factor) out of the modified blackbody spectrum.  In terms of $r=R/R_{id}$, and $T_*$ the power-law energy spectrum, $F_{E}$ is given by:
\begin{equation}
\begin{split}
F_{E} = &\dfrac{1}{4\pi \dsys{2}}  \mathcal{P}_{\tau} \int_{0}^{\nu}\!\!d\nu_s \int_{1}^{40}\!\!dr\, h\,G_{\nu}(x,x_s,T_e,\mathcal{P}_{\tau})\dfrac{1}{\fcol{4}}
		  \dfrac{2 h \nu_s}{c^2} \dfrac{2\pi r /(1+0.25 h_c^2/(r-1)^2)^{1/2}}{\exp\left(h\nu_s (r^{3/4}(1-r^{-1/2})^{-1/4})/k_B T_* \right)-1}\\.
\end{split}
\end{equation}
Note that we have assumed the corona emission is largely isotropic, but we have included the projection factor for blackbody emission from each annulus to half the height of the corona at $r=1$, and we confirmed that $r=40$ is a numerically acceptable cutoff.  Fixing $\Gamma_{\mbox{\scriptsize pli}} = 2.34$ and $T_e=150\mbox{keV}$ ($T_e$ dependence is relatively weak for $T_e \gg \max( T_*, \mbox{25 keV}/\kboltz{})$) we tabulated integrated flux for a range of $T_{\mbox{\scriptsize max}}$ and $\innrMcdot{}$ to be inverted numerically, ultimately obtaining the $\innrMcdot{}$ points in the upper panel of fig.\ref{fig:MdotsData}.  For relatively high $\Tbb{}$ and low $\tau$ they agree fairly well with the simpler method, with the main difference due to the less step-like $\innrMcdot{}$--$\tau$ relationship in the more detailed $P_{\tau}$.  However, to explain observations with higher $F_{pl}$ and lower $\Tbb{}$, the formula quickly requires excessively high $\innrMcdot{}$, and at these implied optical depths the formula itself becomes unreliable.

\end{appendix}

\bibliographystyle{apj}
\bibliography{Paper1f}

\begin{thebibliography}{60}
\expandafter\ifx\csname natexlab\endcsname\relax\def\natexlab#1{#1}\fi

\bibitem[{{Armitage} \& {Livio}(1998)}]{ArmLiv98}
{Armitage}, P.~J., \& {Livio}, M. 1998, {ApJ}, 493, 898

\bibitem[{{Arnaud}(1996)}]{Arn96}
{Arnaud}, K.~A. 1996, in Astronomical Society of the Pacific Conference Series,
  Vol. 101, Astronomical Data Analysis Software and Systems V, ed. G.~H.
  {Jacoby} \& J.~{Barnes}, 17

\bibitem[{{Beckwith} {et~al.}(2008){Beckwith}, {Hawley}, \&
  {Krolik}}]{BeHaKr_2008_RinnrRad}
{Beckwith}, K., {Hawley}, J.~F., \& {Krolik}, J.~H. 2008, {MNRAS}, 390, 21

\bibitem[{{Begelman} {et~al.}(1983){Begelman}, {McKee}, \&
  {Shields}}]{BeMckSh83}
{Begelman}, M.~C., {McKee}, C.~F., \& {Shields}, G.~A. 1983, {ApJ}, 271, 70

\bibitem[{{Bradley} \& {Frank}(2009)}]{BradFrank2009}
{Bradley}, C.~K., \& {Frank}, J. 2009, ApJ, 704, 25

\bibitem[{Cannizzo(1998)}]{Cannizzo_98_IoniInstCycles1}
Cannizzo, J. 1998, ApJ, 494, 366

\bibitem[{{Cao}(2011)}]{Cao_2011_AdafBfields}
{Cao}, X. 2011, {ApJ}, 737, 94

\bibitem[{{Casse} \& {Ferreira}(2000)}]{CasseFerreira_2000_ColdWindCondsIV}
{Casse}, F., \& {Ferreira}, J. 2000, \aap, 353, 1115

\bibitem[{{Chakrabarti} \&
  {Titarchuk}(1995)}]{ChakrabartiTitarchuk_1995_TwoComp}
{Chakrabarti}, S., \& {Titarchuk}, L.~G. 1995, \apj, 455, 623

\bibitem[{{Coriat} {et~al.}(2012){Coriat}, {Fender}, \&
  {Dubus}}]{Coriat_etal_12_PersistentVTransients}
{Coriat}, M., {Fender}, R.~P., \& {Dubus}, G. 2012, \mnras, 424, 1991

\bibitem[{{de Guiran} \& {Ferreira}(2011)}]{deGuiranFerreira_2011_arXiv_Bdrag}
{de Guiran}, R., \& {Ferreira}, J. 2011, ArXiv e-prints, arXiv:1112.5343

\bibitem[{{Done} {et~al.}(2007){Done}, {Gierli{\'n}ski}, \&
  {Kubota}}]{DoneGierKub_07Rev}
{Done}, C., {Gierli{\'n}ski}, M., \& {Kubota}, A. 2007, \aapr, 15, 1

\bibitem[{{Dubus} {et~al.}(1999){Dubus}, {Lasota}, {Hameury}, \&
  {Charles}}]{Dubus_etal_1999_irrPHIIa}
{Dubus}, G., {Lasota}, J.-P., {Hameury}, J.-M., \& {Charles}, P. 1999, \mnras,
  303, 139

\bibitem[{{Ebisuzaki} {et~al.}(1984){Ebisuzaki}, {Sugimoto}, \&
  {Hanawa}}]{EbHaSu84}
{Ebisuzaki}, T., {Sugimoto}, D., \& {Hanawa}, T. 1984, \pasj, 36, 551

\bibitem[{{Esin} {et~al.}(1997){Esin}, {McClintock}, \&
  {Narayan}}]{Esin_etal_1997_TwoComp}
{Esin}, A.~A., {McClintock}, J.~E., \& {Narayan}, R. 1997, \apj, 489, 865

\bibitem[{{Fender} {et~al.}(1999){Fender}, {Corbel}, {Tzioumis}, {McIntyre},
  {Campbell-Wilson}, {Nowak}, {Sood}, {Hunstead}, {Harmon}, {Durouchoux}, \&
  {Heindl}}]{Fender_etal_1999_GX339_jetQuench}
{Fender}, R., {Corbel}, S., {Tzioumis}, T., {et~al.} 1999, {ApJL}, 519, L165

\bibitem[{{Fender} {et~al.}(1998){Fender}, {Southwell}, \&
  {Tzioumis}}]{FenderSoTz_1998_jetlimits}
{Fender}, R.~P., {Southwell}, K., \& {Tzioumis}, A.~K. 1998, {MNRAS}, 298, 692

\bibitem[{{Foulkes} {et~al.}(2010){Foulkes}, {Haswell}, \&
  {Murray}}]{Foulkes_etal_2010_IrradWarping}
{Foulkes}, S.~B., {Haswell}, C.~A., \& {Murray}, J.~R. 2010, \mnras, 401, 1275

\bibitem[{Frank {et~al.}(2002)Frank, King, \& Raine}]{FKR_book02}
Frank, J., King, A., \& Raine, D. 2002, Accretion Power in Astrophysics, 3rd
  edn. (Cambridge)

\bibitem[{Galeev {et~al.}(1979)Galeev, Rosner, \&
  Vaiana}]{Galeev_etal_79_diskBfields}
Galeev, A., Rosner, R., \& Vaiana, G. 1979, {ApJ}, 229, 318

\bibitem[{{Garain} {et~al.}(2012){Garain}, {Ghosh}, \&
  {Chakrabarti}}]{Garain_etal_2012_2FlowMonteCarlo1}
{Garain}, S.~K., {Ghosh}, H., \& {Chakrabarti}, S.~K. 2012, \apj, 758, 114

\bibitem[{{Hessman}(1999)}]{Hessman99}
{Hessman}, F.~V. 1999, {ApJ}, 510, 867

\bibitem[{{Homan} \& {Belloni}(2005)}]{HomanBelloni_05_qHystRev}
{Homan}, J., \& {Belloni}, T. 2005, \apss, 300, 107

\bibitem[{{Hua} \& {Titarchuk}(1995)}]{HuaTitarchuk_1995_iC}
{Hua}, X.-M., \& {Titarchuk}, L. 1995, {ApJ}, 449, 188

\bibitem[{{Janiuk} \& {Czerny}(2011)}]{JaniukCzerny_2011}
{Janiuk}, A., \& {Czerny}, B. 2011, \mnras, 414, 2186

\bibitem[{{Kim} {et~al.}(1999){Kim}, {Wheeler}, \&
  {Mineshige}}]{Kim_etal_1999_TimeDepPHII}
{Kim}, S.-W., {Wheeler}, J.~C., \& {Mineshige}, S. 1999, \pasj, 51, 393

\bibitem[{{King} {et~al.}(2012){King}, {Miller}, {Raymond}, {Fabian},
  {Reynolds}, {Kallman}, {Maitra}, {Cackett}, \&
  {Rupen}}]{King_etal_2012_BWindsInJ17091}
{King}, A.~L., {Miller}, J.~M., {Raymond}, J., {et~al.} 2012, ApJL, 746, L20

\bibitem[{{King} {et~al.}(1997){King}, {Kolb}, \&
  {Szuszkiewicz}}]{King_etal_1997_quickIrrLimits}
{King}, A.~R., {Kolb}, U., \& {Szuszkiewicz}, E. 1997, \apj, 488, 89

\bibitem[{{Lasota}(2001)}]{Lasota_2001_IoniInstRev}
{Lasota}, J.-P. 2001, {NAR}, 45, 449

\bibitem[{{Liu} {et~al.}(2007){Liu}, {Taam}, {Meyer-Hofmeister}, \&
  {Meyer}}]{LiTaMe_HoMe07}
{Liu}, B.~F., {Taam}, R.~E., {Meyer-Hofmeister}, E., \& {Meyer}, F. 2007,
  {ApJ}, 671, 695

\bibitem[{Lubow \& Shu(1975)}]{LubShu75}
Lubow, S., \& Shu, F. 1975, {ApJ}, 198, 383

\bibitem[{{Mayer} \& {Pringle}(2007)}]{MaPr07}
{Mayer}, M., \& {Pringle}, J.~E. 2007, {MNRAS}, 376, 435

\bibitem[{{Meyer} {et~al.}(2007){Meyer}, {Liu}, \&
  {Meyer-Hofmeister}}]{MeLi_MH_07_RecondAndGap}
{Meyer}, F., {Liu}, B.~F., \& {Meyer-Hofmeister}, E. 2007, AAP, 463, 1

\bibitem[{{Meyer} \& {Meyer-Hofmeister}(1983)}]{MeMe_Ho83}
{Meyer}, F., \& {Meyer-Hofmeister}, E. 1983, \aap, 121, 29

\bibitem[{{Meyer-Hofmeister} {et~al.}(2009){Meyer-Hofmeister}, {Liu}, \&
  {Meyer}}]{Me_HoLiMe09}
{Meyer-Hofmeister}, E., {Liu}, B.~F., \& {Meyer}, F. 2009, A\&A, 508, 329

\bibitem[{Montgomery \& Martin(2010)}]{MontgomeryMartin_2010_Lift}
Montgomery, M., \& Martin, E. 2010, ApJ, 722, 989

\bibitem[{{Nandi} {et~al.}(2012){Nandi}, {Debnath}, {Mandal}, \&
  {Chakrabarti}}]{Nandi_etal_2012_TwoFlow}
{Nandi}, A., {Debnath}, D., {Mandal}, S., \& {Chakrabarti}, S.~K. 2012, \aap,
  542, A56

\bibitem[{Narayan \& Yi(1995{\natexlab{a}})}]{NaYi95a}
Narayan, R., \& Yi, I. 1995{\natexlab{a}}, {ApJ}, 444, 231

\bibitem[{Narayan \& Yi(1995{\natexlab{b}})}]{NaYi95b}
---. 1995{\natexlab{b}}, {ApJ}, 452, 710

\bibitem[{{Ogilvie} \& {Dubus}(2001)}]{OgDu01}
{Ogilvie}, G.~I., \& {Dubus}, G. 2001, {MNRAS}, 320, 485

\bibitem[{{Page} {et~al.}(2003){Page}, {Soria}, {Wu}, {Mason}, {Cordova}, \&
  {Priedhorsky}}]{Page_etal_2003_LMCX3obs}
{Page}, M.~J., {Soria}, R., {Wu}, K., {et~al.} 2003, {MNRAS}, 345, 639

\bibitem[{{Pringle}(1992)}]{Pringle_1992_BasicWarping}
{Pringle}, J.~E. 1992, \mnras, 258, 811

\bibitem[{{Proga}(2003)}]{Proga_2003_MHDwinds}
{Proga}, D. 2003, {ApJ}, 585, 406

\bibitem[{{Shafee} {et~al.}(2008){Shafee}, {Narayan}, \&
  {McClintock}}]{ShNaMC_2008_RinnrRadToo}
{Shafee}, R., {Narayan}, R., \& {McClintock}, J.~E. 2008, {ApJ}, 676, 549

\bibitem[{{Shakura} \& {Sunyaev}(1973)}]{ShakSun73}
{Shakura}, N.~I., \& {Sunyaev}, R.~A. 1973, \aap, 24, 337

\bibitem[{{Shields} {et~al.}(1986){Shields}, {McKee}, {Lin}, \&
  {Begelman}}]{ShMckLiBe_1986_CHW2}
{Shields}, G.~A., {McKee}, C.~F., {Lin}, D.~N.~C., \& {Begelman}, M.~C. 1986,
  {ApJ}, 306, 90

\bibitem[{Shimura \& Takahara(1995)}]{ShiTak95}
Shimura, T., \& Takahara, F. 1995, {ApJ}, 445, 780

\bibitem[{{Smale} \& {Boyd}(2012)}]{SmaleBoyd_2012_arXiv}
{Smale}, A.~P., \& {Boyd}, P.~T. 2012, \apj, 756, 146

\bibitem[{{Smith} {et~al.}(2007){Smith}, {Dawson}, \& {Swank}}]{SmDaSw07}
{Smith}, D.~M., {Dawson}, D.~M., \& {Swank}, J.~H. 2007, {ApJ}, 669, 1138

\bibitem[{{Smith} {et~al.}(2002){Smith}, {Heindl}, \& {Swank}}]{SmHeSw02}
{Smith}, D.~M., {Heindl}, W.~A., \& {Swank}, J.~H. 2002, {ApJ}, 569, 362

\bibitem[{{Soria} {et~al.}(2001){Soria}, {Wu}, {Page}, \&
  {Sakelliou}}]{Soria_etal2001}
{Soria}, R., {Wu}, K., {Page}, M.~J., \& {Sakelliou}, I. 2001, A\&A, 365, L273

\bibitem[{{Val-Baker} {et~al.}(2007){Val-Baker}, {Norton}, \&
  {Negueruela}}]{Va_BaNoNe_2007_LMCX3SysProps}
{Val-Baker}, A.~K.~F., {Norton}, A.~J., \& {Negueruela}, I. 2007, in American
  Institute of Physics Conference Series, Vol. 924, The Multicolored Landscape
  of Compact Objects and Their Explosive Origins, ed. T.~{di Salvo}, G.~L.
  {Israel}, L.~{Piersant}, L.~{Burderi}, G.~{Matt}, A.~{Tornambe}, \& M.~T.
  {Menna}, 530--533

\bibitem[{{van der Klis} {et~al.}(1985){van der Klis}, {Clausen}, {Jensen},
  {Tjemkes}, \& {van Paradijs}}]{vdK_etal_1985}
{van der Klis}, M., {Clausen}, J.~V., {Jensen}, K., {Tjemkes}, S., \& {van
  Paradijs}, J. 1985, \aap, 151, 322

\bibitem[{{Viallet} \& {Hameury}(2007)}]{ViHa_2007_2dSRM}
{Viallet}, M., \& {Hameury}, J.-M. 2007, A\& A, 475, 597

\bibitem[{{Wang} {et~al.}(2012){Wang}, {Cheng}, \&
  {Li}}]{WaChLi_2012_ClumpyAccr}
{Wang}, J.-M., {Cheng}, C., \& {Li}, Y.-R. 2012, {ApJ}, 748, 147

\bibitem[{{Wilms} {et~al.}(2001){Wilms}, {Nowak}, {Pottschmidt}, {Heindl},
  {Dove}, \& {Begelman}}]{Wilms_etal_2001_LMCXsObs}
{Wilms}, J., {Nowak}, M.~A., {Pottschmidt}, K., {et~al.} 2001, \mnras, 320, 327

\bibitem[{{Woods} {et~al.}(1996){Woods}, {Klein}, {Castor}, {McKee}, \&
  {Bell}}]{WoodsEtAl96}
{Woods}, D.~T., {Klein}, R.~I., {Castor}, J.~I., {McKee}, C.~F., \& {Bell},
  J.~B. 1996, {ApJ}, 461, 767

\bibitem[{{Zanni} {et~al.}(2007){Zanni}, {Ferrari}, {Rosner}, {Bodo}, \&
  {Massaglia}}]{Zanni_etal_2007_specJetSims}
{Zanni}, C., {Ferrari}, A., {Rosner}, R., {Bodo}, G., \& {Massaglia}, S. 2007,
  \aap, 469, 811

\bibitem[{{Zdziarski} {et~al.}(1994){Zdziarski}, {Fabian}, {Nandra}, {Celotti},
  {Rees}, {Done}, {Coppi}, \& {Madejski}}]{Zdz_etal_1994_iC_app}
{Zdziarski}, A.~A., {Fabian}, A.~C., {Nandra}, K., {et~al.} 1994, \mnras, 269,
  L55

\bibitem[{{Zimmerman} {et~al.}(2005){Zimmerman}, {Narayan}, {McClintock}, \&
  {Miller}}]{Zimmerman_etal_2005_bbSpec}
{Zimmerman}, E.~R., {Narayan}, R., {McClintock}, J.~E., \& {Miller}, J.~M.
  2005, \apj, 618, 832

\end{thebibliography}

\clearpage

\end{document}